\documentclass[10pt, aps, prd,
reprint, % for double column
notitlepage, nofootinbib, longbibliography, showpacs, letter,
superscriptaddress,
]{revtex4-1}

\usepackage{pstricks,pst-fractal}
\usepackage{amsmath,amssymb,bm}
\usepackage{longtable}
\usepackage{graphicx}
\usepackage{verbatim}
\usepackage[breaklinks=true,colorlinks=true,urlcolor=blue,linkcolor=red,
    citecolor=red]{hyperref}

\begin{document}

\title{Role of zero point energy in promoting ice formation 
in a spherical drop of water} 

\author{Prachi Parashar}
\email{Prachi.Parashar@jalc.edu}
%\homepage{https://www.ntnu.edu/employees/prachi.parashar}
\affiliation{John A. Logan College,
Carterville, Illinois 62918, USA}
\affiliation{Department of Energy and Process Engineering,
Norwegian University of Science and Technology, NO-7491 Trondheim, Norway}

\author{K. V. Shajesh}
\email{kvshajesh@gmail.com} 
%\homepage{http://www.physics.siu.edu/~shajesh}
\affiliation{Department of Physics, Southern Illinois University--Carbondale,
Carbondale, Illinois 62901, USA}
\affiliation{Department of Energy and Process Engineering,
Norwegian University of Science and Technology, NO-7491 Trondheim, Norway}

\author{Kimball A. Milton}
\email{kmilton@ou.edu} 
%\homepage{http://www.nhn.ou.edu/~milton}
\affiliation{Homer L. Dodge Department of Physics and Astronomy,
University of Oklahoma, Norman, Oklahoma 73019, USA}

\author{Drew F. Parsons}
\email{D.Parsons@murdoch.edu.au}
%\homepage{http://profiles.murdoch.edu.au/myprofile/drew-parsons}
\affiliation{Discipline of Chemistry \& Physics, CSHEE,
Murdoch University, 90 South Street, Murdoch, WA 6150, Australia}

\author{Iver Brevik}
\email{iver.h.brevik@ntnu.no}
%\homepage{https://www.ntnu.edu/employees/iver.h.brevik}
\affiliation{Department of Energy and Process Engineering,
Norwegian University of Science and Technology, NO-7491 Trondheim, Norway}

\author{Mathias Bostr{\"o}m}
\email{mathias.a.bostrom@ntnu.no}
%\homepage{https://www.mn.uio.no/fysikk/english/people/aca/mathbo/index.html}
\affiliation{Department of Energy and Process Engineering,
Norwegian University of Science and Technology, NO-7491 Trondheim, Norway}
\affiliation{Centre for Materials Science and Nanotechnology,
Department of Physics,
University of Oslo, P. O. Box 1048 Blindern, NO-0316 Oslo, Norway}

\date{\today}

%------------------------------------
\begin{abstract}
We demonstrate that the Lifshitz interaction energy (excluding
the self-energies of the inner and outer spherical regions)
for three concentric spherical dielectric media can be evaluated
easily using the immense computation power in recent processors
relative to those of a few decades ago. As a prototype, we compute
the Lifshitz interaction energy for a spherical shell of water
immersed in water vapor of infinite extent while enclosing a
spherical ball of ice inside the shell, such that two concentric
spherical interfaces are formed: one between solid ice and
liquid water and the other between liquid water and gaseous vapor.
We evaluate the Lifshitz interaction energy for the above
configuration at the
triple point of water when the solid, liquid, and gaseous states of
water coexist, and, thus, extend the analysis of Elbaum and Schick
in Phys.~Rev.~Lett.~{\bf 66} (1991) 1713 to spherical configurations.
We find that, when the Lifshitz energy contributes dominantly to the 
total energy of this system, which is often the case when electrostatic 
interactions are absent, a drop of water surrounded by vapor of
infinite extent is not stable at the triple point. This instability,
that is a manifestation of the quantum fluctuations in the medium,
will promote formation 
of ice in water, which will then grow in size indefinitely.
This is a consequence of the finding here that the Lifshitz energy
is minimized for large (micrometer size) radius of the ice ball and
small (nanometer size) thickness of the water shell surrounding the
ice. These results might be relevant to the formation of hail in
thunderclouds. These results are tentative in that the self-energies
are omitted; surface tension and nucleation energy are not considered.
\end{abstract}

\maketitle
%------------------------------------
%\tableofcontents
\newpage
%--------------------------------------------
\section{Introduction}

The term Casimir effect is often used to refer to the entire 
phenomena associated with quantum fluctuations. Other closely related
terminologies are quantum vacuum energy, zero point energy, Lifshitz energy,
London dispersion forces, and van der Waals interactions.
The ideas governing the van der Waals interactions~\cite{Waals:1873sl}
and London dispersion 
forces~\cite{London:1930a,London:1930b,Hettema:2001cq}
originated in attempts to understand
the interactions of neutral, but polarizable, molecules of gases
that deviated in their characteristics from the ideal gas law.
Casimir and Polder~\cite{Casmir:1947hx}
later generalized these calculations to include retardation effects.
The concept of zero point energy on the other hand originated in the
1910's in works of 
Refs.~\cite{Planck:1914rht,Einstein:1910ser,Einstein:1913nab} 
where the focus was to understand the blackbody 
radiation~\cite{Planck:1901sne} in the limit of zero temperature.
A priori it was not expected that the theory of radiation would have 
anything to do with inter-atomic forces.
However, the astonishing feat of Casimir~\cite{Casimir:1948pc}
was in showing that 
London dispersion forces, or the van der Waals interactions,
were manifestations of the zero point energy.

\begin{figure}
\includegraphics{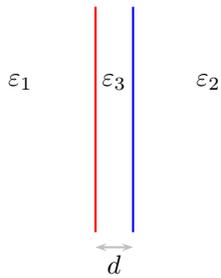}
\caption{Three planar regions described by dielectric functions
$\varepsilon_1$, $\varepsilon_3$, and $\varepsilon_2$,
such that the thickness of the confined medium is $d$. }
\label{fig-planes-123}
\end{figure}

Casimir evaluated the energy of a planar cavity with perfectly conducting
walls, an overly idealized system,
that is obtained from the configuration of Fig.~\ref{fig-planes-123}
when the regions labeled as $\varepsilon_1$ and $\varepsilon_2$ are
perfect electrical conductors that are separated by 
vacuum in the background region
labeled $\varepsilon_3$.
Lifshitz~\cite{Lifshitz:1956sb} generalized Casimir's result
by evaluating the energy for a configuration of Fig.~\ref{fig-planes-123}
consisting of two dielectric media of infinite extent 
separated by vacuum.
The Lifshitz energy leads to the Casimir energy in 
the perfect conducting limit
of the dielectric functions for the outer media.
Dzyaloshinskii, Lifshitz, and Pitaevskii (DLP)~\cite{Dzyaloshinskii:1961fw}
extended these considerations for the case when the background region
in the planar configuration of Fig.~\ref{fig-planes-123}
is another uniform dielectric medium. 
The main idea underlying these groundbreaking works is that quantum
fluctuations of fields in the media can be manifested in physical phenomena
involving dielectrics. Among these, we point out that
the configurations considered by Casimir
and Lifshitz always lead to an attractive pressure (tending to
decrease the thickness of the intervening medium).
In contrast, the configurations considered
by DLP allows for the pressure to be attractive or repulsive.

%----------------
%----------------

Elbaum and Schick~\cite{Elbaum1991wi}, using the result of DLP
in conjunction with the existing data for the dielectric functions
for ice and water, showed that an interface of solid ice 
and gaseous vapor is unstable at the triple point of water.
They showed that quantum fluctuations 
in the electromagnetic fields in the media induces the
formation of a 3.56\,nm thick layer of liquid water at the interface,
intervening between solid ice and gaseous vapor.
The temperature of the triple point of water sets the scale for
the energy of the system and the associated characteristic frequency
obtained by dividing the energy by the Planck constant $\hbar$
corresponds to the frequency of the lowest Matsubara mode,
equal to $2\pi k_BT/\hbar\approx 2.5\times 10^{14}$\,rad/s. 
The source of the instability of the interface predicted by
Elbaum and Schick was associated to the fact that solid
ice is more polarizable than liquid water for frequencies
larger than the 71st Matsubara mode,
$\omega_{c2}\approx 71 \times (2.5\times 10^{14}$\,rad/s), while
solid ice is less polarizable than liquid water for frequencies
smaller than $\omega_{c2}$, and
the static polarizability of solid ice is larger than that of liquid water.
A remarkable feature of the Elbaum-Schick effect in water
is that it necessarily requires taking retardation effects into
consideration, that is, the effect disappears in a non-retarded analysis. 
This is striking because in planar configurations 
with vacuum as the background medium
the retardation effects become relevant only when the thickness of
the vacuum is hundreds of nanometer thick.
However, if we introduce an intervening medium 
as done in the general configuration of DLP,
it is possible to have retardation effects
play a role at very small distances, the Elbaum-Schick effect
being a classic example. This was already anticipated by DLP
in the context of wetting of a wall.
Recently in Ref.~\cite{Thiyam:2018lsr} this same idea was
exploited to reverse the direction of torque as the separation
distance between two anisotropic dielectric media is changed.
We also verified that the Elbaum-Schick effect
does not get washed away in the weak approximation, 
applicable for dilute dielectric media, which keeps the retardation
effect and drops the higher order terms after expanding the logarithm
in the expression for the Lifshitz energy. 

It is important to emphasize that this effect cannot be thought
of in terms of van der Waals energies, which refer to dilute,
nonretarded limit. The effects we discuss in terms of the
Lifshitz theory depend crucially on non-pairwise-additive
forces and retardation. A Hamaker construction fails to capture
the physics.

In this article we will evaluate the Lifshitz interaction energy,
excluding the self-energies of the inner and outer dielectric regions.
This calls for the definition of Lifshitz energy and some clarification
on what we are not achieving in our calculations. To this end,
we point out, though it has been surely known all along, that
the energy for the configuration of Fig.~\ref{fig-planes-123}
allows the decomposition~\cite{Kenneth:2006vr,Shajesh:2011ef}:
\begin{equation}
E = E_3 +\Delta E_1 +\Delta E_2 + E_{12}.
\label{edecom}
%Eq.\,(\ref{edecom})
\end{equation}
Here $E$ is the total energy; $E_3$ is the total energy
when both interfaces are moved infinitely far away from
each other to infinity, such that all space is filled
with medium $\varepsilon_3$; $\Delta E_1 =E_1-E_3$
and $\Delta E_2=E_2-E_3$ are self-energies 
required to create systems with single
interfaces when the other interface is moved to 
$\pm\infty$ respectively; $E_{12}$ is
the interaction energy between media
$\varepsilon_1$ and $\varepsilon_2$.
The interaction energy $E_{12}$
is the only contribution to the total energy
that depends on the position and orientation of both media 
and determines the forces between them.
This decomposition is generic,
irrespective of $\varepsilon_3$ being vacuum or another medium.
The importance of the decomposition of energy in Eq.\,(\ref{edecom})
is the fact that the interaction energy $E_{12}$ is unambiguously finite
by construction if media $\varepsilon_1$ and $\varepsilon_2$ are disjoint,
even while the self-energies
$\Delta E_1$, $\Delta E_2$, and $E_3$ remain divergent.
Of considerable importance is the fact that self-energies 
may include the surface energies leading to surface tensions
in the interfaces, however, due to the lack of predictive power
in the face of divergences we will not discuss
these terms in this article. The interaction
energy $E_{12}$ is called the Lifshitz energy and this part of energy
will be the subject matter of this article.
The lack of a complete understanding to date of the divergent expressions
in energy and omission of the associated contributions 
to energy all together here
will remain a limitation of our analysis here.

Elbaum and Schick's conclusion that quantum fluctuations induce
the formation of a thin layer of liquid water at the interface
of solid ice and gaseous vapor is valid for planar configurations.
It is, then, of interest to inquire
if these considerations change for curved geometries.
In this article we extend the analysis of Elbaum and Schick
to spherical concentric interfaces of solid ice, liquid water,
and water vapor. We conclude that a spherical drop of water immersed
inside gaseous vapor of infinite extent is unstable at the triple point
of water. Quantum fluctuations promote formation of solid ice inside
the drop of liquid water, which will then grow in size indefinitely.
Once the solid ice has grown sufficiently large its surface can be
approximated to that of a plane and in this limit the results of the
planar configuration apply and the liquid water attains a thickness of
3.56\,nm at equilibrium. The phenomena of quantum fluctuations
promoting formation of ice in water, to our knowledge,
has not been reported or mentioned in the literature before.
This is expected to prompt a plethora of applications and studies
associated to this phenomena, a few of which we mention in the 
last section and hope to explore in future publications.

A caveat must be stressed: We ignore the self-energies of each
phase, as well as the associated surface tensions. We also realize
that the energies involved here are very small compared to the
nucleation energies. Surface tensions for the two interfaces
involving water here are of order $10^{-1}$\,J/m$^2$,
while the energies resulting from the Lifshitz effects
we consider are of order $10^{-7}$\,J/m$^2$. So, we are 
considering quite small, but, we believe, significant effects.

Even though the expressions for Lifshitz energy 
(Helmholtz free energy) reported in
this article are sufficiently general, we will consistently
use solid ice for region 1 described by $\varepsilon_1$,
liquid water for the background region $\varepsilon_3$,
and gaseous water vapor for region 2 described by $\varepsilon_2$. 
The discussions in this article will
be confined to the temperature and pressure associated with
the triple point of water, $273.16\,$K and 611.657\,Pa,
when solid ice, liquid water, and water vapor, can coexist.

It should be emphasized that we are considering the quantum
electrodynamic Casimir effect, due to the electrical properties
of the materials, and not the thermodynamic critical Casimir
effect~\cite{Krech:1994},
which is of quite a different character, and should not be
relevant at the triple point, which is far from the critical
point of water ($T_c=647$\,K, $p_c=22$\,MPa).
At a critical point, as opposed to a triple point, the
associated correlation length becomes infinite. 
Similarly, we have not accounted for the plausible fluctuations
in surface imperfections. Thus, our
discussion here is applicable for zero correlation lengths
in the associated fluctuations.

%--------------------------------------------
\section{Elbaum-Schick effect}
\label{sec-EB-planar}

The Lifshitz interaction energy per unit area for the planar
configuration of Fig.~\ref{fig-planes-123},
consisting of three dielectric media
with negligible magnetic permeabilities, $\mu_i=1$,
such that the sandwiched medium has thickness $d$, is given by,
\begin{eqnarray}
{\cal E}(d) &=& \frac{\hbar c}{4\pi^2 a_0}
{\sum_{n=0}^\infty}{^\prime} \int_0^\infty kdk 
\nonumber \\ && \times
\ln \left[ 1 - r_{31}^E r_{32}^E e^{-2\kappa_3 d} \right]
\left[ 1 - r_{31}^H r_{32}^H e^{-2\kappa_3 d} \right], \hspace{3mm}
\label{def-Life}
\end{eqnarray}
where the reflection coefficients for the transverse electric
($E$) and transverse magnetic ($H$) modes are given by
\begin{equation}
r_{ij}^E = - \left( \frac{\kappa_i-\kappa_j}{\kappa_i+\kappa_j} \right), \quad 
r_{ij}^H = - \left( \frac{\varepsilon_j\kappa_i-\varepsilon_i\kappa_j}
     {\varepsilon_j\kappa_i+\varepsilon_i\kappa_j} \right), 
\label{def-refcoe}
\end{equation}
respectively, in terms of the effective refractive index
\begin{equation}
\kappa_i = \sqrt{k^2 + \frac{n^2}{a_0^2} \varepsilon_i}, \qquad i=1,3,2.
\label{def-kap}
\end{equation}
The prime on the summation symbol in Eq.\,(\ref{def-Life}) indicates
that the $n=0$ term is to be multiplied by a factor of $1/2$.
We have defined the constant
\begin{equation}
a_0 = \frac{\hbar c}{2\pi k_BT}
\label{natscd}
\end{equation}
with dimensions of length,
which introduces a natural scale for distance in the discussion.
The corresponding scale for energy is set by the coefficient
$\hbar c/(4\pi^2 a_0) =k_BT/(2\pi)$ in Eq.\,(\ref{def-Life}).
For typical dielectric materials at room temperature this distance
$a_0$ is in the micrometer range corresponding to an energy in milli
electron-volts.
At the triple point of water the distance $a_0$ in Eq.\,(\ref{natscd}) 
is evaluated to be 
\begin{equation}
a_0=1.3342\,\mu\text{m}.
\end{equation}
and the energy $\hbar c/(4\pi^2 a_0)=k_BT/(2\pi)$
in Eq.\,(\ref{def-Life}) equals 3.7463\,meV.
% at the triple point of water.

%------------------------------------
\subsection{Dielectric function}

%----------------
\begin{figure}
\includegraphics{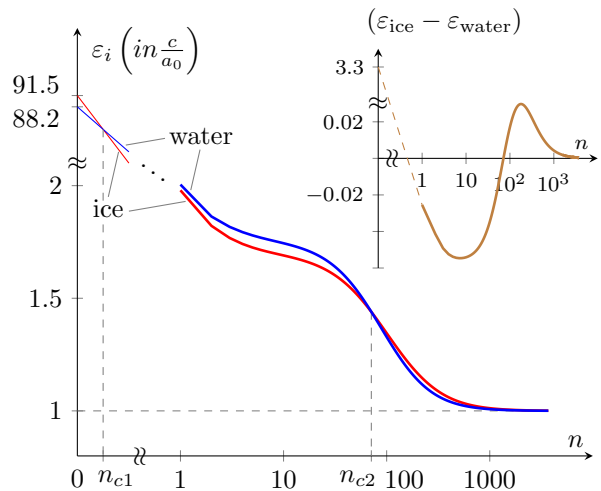}
\caption{
Dielectric functions of solid ice and liquid water, obtained using the
fitting parameters used by Elbaum and Schick in Ref.~\cite{Elbaum1991wi},
plotted with respect to the Matsubara mode number
$n$ ranging from 1 to 3700. The
discrete frequency is $nc/a_0$, where $c/a_0 =2.2470\times 10^{14}$\,rad/s.
The dielectric function at zero frequency ($n=0$) is huge 
and only sketched out for illustration, because it is hard to capture
it on the same scale. The dielectric functions of solid ice and liquid water
when extrapolated as a smooth line, even though they are actually
discrete points,
intersect at two points, first at $n_{c1}$ between $n=0$ and $n=1$,
and then again at $n_{c2}$ between $n=71$ and $n=72$. The difference
in the dielectric functions of solid ice and liquid
water is plotted in the inset, where, again, the $n=0$ contribution
is illustrated as a cartoon.
}
\label{fig-diel-versus-omega}
\end{figure}%
%----------------

The permittivities $\varepsilon_i$ in Eqs.\,(\ref{def-refcoe})
and (\ref{def-kap}) are functions of the
discrete imaginary frequency, the Matsubara frequency, $inc/a_0$,
\begin{equation}
\varepsilon_j =\varepsilon_j\left(i n\frac{c}{a_0} \right), \qquad j=1,3,2.
\label{diel-depD}
\end{equation}
%Here the $i$ inside the parenthesis is $\sqrt{-1}$, while the $i$'s
%in the subscripts is the label for the respective
%regions in Fig.~\ref{fig-planes-123}.
The dielectric functions for $j=1,3,2,$ for solid ice, liquid water,
and gaseous water vapor, respectively,
are generated using the damped oscillator model
for the dielectric response, following Elbaum and Schick~\cite{Elbaum1991wi},
\begin{equation}
\varepsilon(\omega) = 1+
\sum_j \frac{f_j}{e_j^2 -i\hbar\omega\, g_j - (\hbar\omega)^2},
\label{diel-res-d}
\end{equation}
where $e_j$, $f_j$, and $g_j$ are given by the values
listed in Table~1 of Ref.~\cite{Elbaum1991wi}.
The dielectric response at zero frequency for solid ice and liquid water are
\begin{subequations}
\begin{eqnarray}
\varepsilon_\text{ice}(0) &=& 91.5, \\
\varepsilon_\text{water}(0) &=& 88.2,
\end{eqnarray}
\end{subequations}
respectively.
Data for dielectric functions were generated for $n$ spanning 0 to 3700,
which were sufficient for convergence of the Lifshitz interaction energy
in the regime of interest. The plots of these dielectric functions 
as a function of the Matsubara mode number $n$ are presented in 
Fig.~\ref{fig-diel-versus-omega}.
The dielectric function at zero frequency ($n=0$) for both
ice and water is huge and could not be captured on the same scale,
but, we sketched the intersection as a cartoon to illustrate the point.
These plots, for solid ice and liquid water, intersect at two points,
first at $n_{c1}$ between $n=0$ and $n=1$, and then again at $n_{c2}$
between $n=71$ and $n=72$. The difference in the dielectric functions
of solid ice and liquid water, which plays a central role in our
discussion, is plotted in the inset. Note, in particular, how the $n=0$
contribution for ice and water dwarfs the contribution from non-zero values.

%------------------------------------
\subsection{Model dependence of dielectric functions}

The results in this article are dependent on the intersections
in Fig.~\ref{fig-diel-versus-omega}.
It is, then, of significance
to investigate and quantify the sensitivity of the effect,
discussed by Elbaum and Schick in Ref.\,\cite{Elbaum1991wi}
and by us in this article, on the fitting parameters for
the dielectric functions of solid ice and liquid water. 
Some of us with other collaborators have taken up this
investigation separately, and thus it will have to be
reported at a different venue. We content the reader by stating that the
results seem to depend on the models. 
The results are sensitive to small changes in the dielectric functions
for water and ice, originating from impurities, salts, or
from improved models that incorporate a larger range
of optical data~\cite{Fiedler:2019ddw}.
This is not a limitation
for this article, because the purpose of this article is to demonstrate
that the Lifshitz interaction energy for concentric dielectric
media can be evaluated easily, and the actual example we work out
is only a prototype. However, our final results specific to the
configuration of ice, water, and vapor, are indeed sensitive to
the model parameters of the dielectric functions and thus
are at best provisional.

%------------------------------------
\subsection{Lifshitz energy for planar geometry}

\begin{figure}
\includegraphics[width=8.5cm]{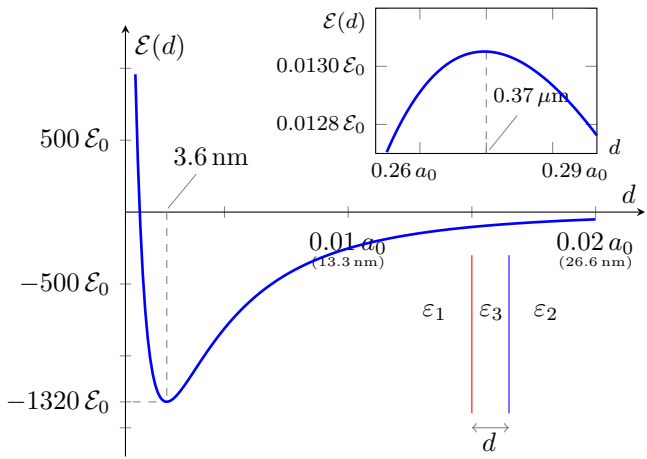}
\caption{Lifshitz energy per unit area ${\cal E}(d)$
for liquid water of thickness $d$ sandwiched between solid ice
and gaseous water vapor plotted as a function
of $d$. The thickness $d$ is marked in units of
$a_0=\hbar c/(2\pi k_BT) \approx 1.3342\,\mu$m, and
the Lifshitz energy per unit area is marked in units of
${\cal E}_0=\hbar c/(4\pi^2 a_0^3)=k_BT/(2\pi)/a_0^2
\approx 3.3720 \times 10^{-10}$\,J/m$^2
\approx 2.1046 \times 10^{-11}$\,eV/\AA$^2$.
The Lifshitz energy has a minima at $d=3.56$\,nm.
The Lifshitz energy also has a maxima at $d=0.37\,\mu$m, which
is shown in the inset-plot. The Lifshitz energy tends to
zero from the positive side for large thickness,
and goes to positive infinity for zero thickness.
}
\label{fig-ES-Stability-123}
\end{figure}

Using the model for the dielectric response in Eq.\,(\ref{diel-res-d})
for the fitting parameters used by Elbaum and Schick~\cite{Elbaum1991wi},
plotted in Fig.~\ref{fig-diel-versus-omega},
the Lifshitz energy per unit area as a function of thickness $d$
of liquid water layer is plotted in Fig.\,\ref{fig-ES-Stability-123}.
The Lifshitz energy diverges to positive infinity for zero thickness $d$
of liquid water, implying an instability of such an interface.
That is, quantum fluctuations will induce the
formation of a thin layer of liquid water at the
interface of solid ice and gaseous vapor. The 
Lifshitz energy associated with two bodies, say two dielectric media
separated by vacuum, diverges to negative infinity when the two media
come in contact. Additionally, the Lifshitz energy goes to (positive) zero
for large thickness. For intermediate distances the Lifshitz energy
has a negative
minimum for $d\sim 0.00267 a_0 \approx 3.56\,$nm. The existence of this
minimum implies that at the triple point of water
it is energetically favorable to form a layer of water
at the interface of ice and vapor. In other words, an interface of
solid ice and gaseous vapor is highly unstable because of the
positive infinite energy associated 
with zero thickness of water in Fig.\,\ref{fig-ES-Stability-123}.
At equilibrium the thickness of water formed at the interface is 3.56\,nm. 

The Lifshitz energy has a local maximum (positive) value
of $0.0131\,{\cal E}_0$ when the thickness of water layer is 
$d\sim 0.275 a_0\approx 0.37\,\mu\text{m}$,
which is shown in the inset of Fig.\,\ref{fig-ES-Stability-123}.
Thus, it is implied that the Lifshitz energy approaches zero
from positive values of energy for large thickness of the water layer.
This is consistent with the fact that at large distances the Lifshitz
energy is completely characterized by the $n=0$ contribution.
This observation, in principle, implies that complete melting of ice
is possible if the water layer is thicker than $0.37\,\mu$m initially.
However, the Lifshitz energy peaks here with a very tiny positive
value of $0.0130\,{\cal E}_0$, which is very small relative to
${\cal E}_0 =k_BT/(2\pi)/a_0^2$.
Thus, complete melting is unlikely due to disturbances
in energy from the surroundings.

%----------------
%----------------

Thus the Lifshitz energy has two extrema, a minimum at $d=3.56$\,nm,
and a maximum at $d=0.37\,\mu$m. These extremum points are roughly
numerically estimated in terms of the two intersection points,
$n_{c1}\sim 0.99$ and $n_{c2}=71$, 
in the plots of the dielectric functions of solid ice and liquid water in
Fig.~\ref{fig-diel-versus-omega}. We crudely estimated
$n_{c1}\sim 0.99$ by assuming a linear interpolation
between the data points at $n=0$ and $n=1$ in Fig.~\ref{fig-diel-versus-omega}.
The two intersection points in Fig.~\ref{fig-diel-versus-omega}
correspond to frequencies
\begin{equation}
\omega_{ci} = n_{ci} \frac{c}{a_0}, \qquad i=1,2,
\end{equation}
which leads to
\begin{subequations}
\begin{eqnarray}
\omega_{c1} &=& 1.80\times 10^{14}\,\text{rad/s}, \\
\omega_{c2} &=& 1.60\times 10^{16}\,\text{rad/s}.
\end{eqnarray}
\end{subequations}
In terms of these critical frequencies a rough numerical estimate
of the extremum values for the thickness of water layer is obtained
using~\cite{Bostrom:2017isw} 
\begin{equation}
d_i \sim \frac{c}{\omega_{ci}}
\frac{1}{2\sqrt{\varepsilon_3(i\omega_{ci})}}, \qquad i=1,2.
\label{dmin-esP}
\end{equation}
This expression leads to $d_1\approx 0.41\,\mu$m and $d_2\approx 7.83$nm,
which are in the right ballpark of $0.37\,\mu$m and $3.56$nm, respectively.
We have been unable to find a more accurate
analytical estimate, because we are dealing
with a discrete function of the Matsubara mode numbers $n$,
in addition to the fact that
the zero mode behaves significantly differently from other
modes~\cite{Bostrom:2017isw}. 

%------------------------------------
\subsection{Incomplete surface melting}
\label{sec-inc-sur-mel}

Melting of a solid into liquid at melting point $T_m$
is typically explained as a phase transition in thermodynamics.
Another explanation proposed by Weyl in 1951~\cite{Weyl:1951is}
and theorized by Fletcher in 1968~\cite{Fletcher:1968is} rests on a 
microscopic theory in which onset of melting happens
at temperatures less than $T_m$. The proposal is that the
energy of the ice or water surface is lowered when the dipole
moments of the water molecules orient in assembly. This leads
to the formation of an electric double layer at surfaces
of water and ice. The electrostatic interactions of
such surfaces, neglecting dispersion forces completely,
led to a power law behavior of $d\sim t^{-1/3}$,
where $t=1-(T/T_m)$, $T<T_m$,
for the thickness $d$ of liquid water formed on the surface
of ice at temperatures slightly below the melting point.
Thus, as the temperature approaches the melting point
a thin layer of liquid water is formed at the surface
which then grows to infinite thickness as the temperature
approaches the melting point. 
These conclusions remain mostly the same even when non-retarded
dispersion interactions are taken in account.
This is called (complete) surface melting and seems to be
a well studied microscopic explanation of melting.
However, data from different experiments are not in concord with 
the specific power law behavior mentioned
above~\cite{Dash:1995cei,Elbaum:1995wcp,Slater:2019ms}.

The implication of Elbaum and Schick's results in
Ref.~\cite{Elbaum1991wi} is that the surface melting
for ice is incomplete.
That is, the thickness of water layer remains finite as the 
temperature approaches the melting point.
It was hard to confirm this accurately in the 
experiment by Elbaum et al.~\cite{Elbaum:1993ims}.
The challenge seems to be with determining
the triple point of water precisely,
and the formation of patches of water drops~\cite{Elbaum:1995wcp}
which probably could be associated with an unevenly flat
surface of ice. We will explore the curvature dependence of
Elbaum and Schick's results in Sec.~\ref{sec-ES-sphere}.
Experimental confirmation of incomplete
surface melting remains open~\cite{Slater:2019ms}.

%------------------------------------
\section{Lifshitz energy for concentric spherical geometry}

\begin{figure}
\includegraphics{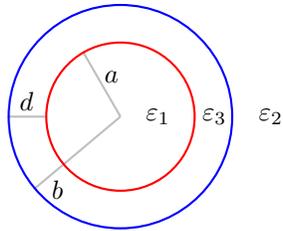}
\caption{Three concentric spherical regions described by dielectric functions
$\varepsilon_1$, $\varepsilon_3$, and $\varepsilon_2$,
separated by the interfaces at radii $a$ and $b=a+d$. }
\label{fig-con-spheres-12}
\end{figure}

The Casimir energy for a perfectly conducting spherical shell was
first calculated by Boyer in 1968,
which surprisingly had the opposite
sign relative to the Casimir energy of two 
parallel plates~\cite{Casimir:1953eqr,Boyer:1968uf}.
The calculation was attempted for a dielectric ball   
in Ref.\,\cite{Milton1980bc}. However, irrespective of the
particular regularization procedures used in the calculation,
the $a_2$ heat kernel coefficient is nonzero, which seems to
suggest that there is no way to make the Casimir energy of a
dielectric ball finite, except for isorefractive cases
($\varepsilon\mu=1$)~\cite{Brevik:1983mcd}. The understanding of this
divergent phenomena associated with a single spherical interface
is generally accepted to be
unsatisfactory \cite{Milton1980bc,Candelas1982sc}.
These calculations evaluated the term $\Delta E_1$
in Eq.\,(\ref{edecom}) for a spherical interface,
with the background region chosen to be a homogeneous medium,
which we pointed out give a divergent contribution.
Here we calculate the interaction energy $E_{12}$ in Eq.\,(\ref{edecom})
for the concentric spherical configuration in Fig.~\ref{fig-con-spheres-12}.
The interaction energy $E_{12}$, by construction, is devoid of divergences
and thus can be evaluated unambiguously.
The concentric spherical configuration of Fig.~\ref{fig-con-spheres-12},
for the case when the inner and outer regions consist of identical
material and the intervening region is vacuum,
$\varepsilon_1=\varepsilon_2$ and $\varepsilon_3=1$,
was studied first, and to our knowledge the only time in literature, by 
Brevik et al.~\cite{Brevik2001am,Brevik2001at,Brevik2002pg}.
However, the numerical estimates reported there were not satisfactory
probably because the necessary convergence was not achievable
with the computational power in the computers of those days.
Our expression for the interaction energy here is a straightforward
generalization
of that in Refs.~\cite{Brevik2001am,Brevik2001at,Brevik2002pg},
obtained by keeping all three regions in 
Fig.~\ref{fig-con-spheres-12} distinct. In addition we
report comprehensive numerical estimates for the interaction energy
for the particular example of ice, water, and vapor,
which can be easily reproduced
for other cases using the methods prescribed here. 
In Ref.~\cite{Parashar:2017sgo}, the interaction energy for
concentric spherical configurations constructed from $\delta$-function
spheres were reported, which is different from the study here.
We emphasize that we are not including the self-energies 
of the interior and exterior regions, which leads to an unknown
systematic error.

%----------------
%----------------

For the spherical geometry of Fig.~\ref{fig-con-spheres-12} with 
interfaces at radii $a$ and $b$ the Lifshitz energy $E(a,b)$ is given by
\cite{Parashar:2017sgo,Shajesh:2017ssa}
\begin{eqnarray}
E(a,b) &=& k_BT
{\sum_{n=0}^\infty}{^\prime} \sum_{l=0}^\infty (2l+1) 
\nonumber \\ && \times 
\ln \left[ 1 - r_{31}^E(a) r_{32}^E(b) \right]
\left[ 1 - r_{31}^H(a) r_{32}^H(b) \right], \hspace{5mm}
\label{enSpL}
\end{eqnarray}
where the various scattering coefficients are given by
\begin{subequations}
\begin{eqnarray}
r_{31}^E(a) &=& \frac{\zeta_3 \, 
\text{i}_l(\zeta_1a) \bar{\text{i}}_l(\zeta_3a) -\zeta_1 \, 
\bar{\text{i}}_l(\zeta_1a) \text{i}_l(\zeta_3a)} { \zeta_3 \, 
\text{i}_l(\zeta_1a) \bar{\text{k}}_l(\zeta_3a) -\zeta_1 \, 
\bar{\text{i}}_l(\zeta_1a) \text{k}_l(\zeta_3a)}, \\
r_{31}^H(a) &=& \frac{\zeta_1 \, 
\text{i}_l(\zeta_1a) \bar{\text{i}}_l(\zeta_3a)
-\zeta_3 \, \bar{\text{i}}_l(\zeta_1a) \text{i}_l(\zeta_3a)}
{ \zeta_1 \, \text{i}_l(\zeta_1a) \bar{\text{k}}_l(\zeta_3a)
-\zeta_3 \, \bar{\text{i}}_l(\zeta_1a) \text{k}_l(\zeta_3a)}, \\
r_{32}^E(b) &=& \frac{\zeta_2 \, 
\text{k}_l(\zeta_3b) \bar{\text{k}}_l(\zeta_2b)
-\zeta_3 \, \bar{\text{k}}_l(\zeta_3b) \text{k}_l(\zeta_2b)}
{ \zeta_2 \, \text{i}_l(\zeta_3b) \bar{\text{k}}_l(\zeta_2b)
-\zeta_3 \, \bar{\text{i}}_l(\zeta_3b) \text{k}_l(\zeta_2b)}, \\
r_{32}^H(b) &=& \frac{\zeta_3 \, 
\text{k}_l(\zeta_3b) \bar{\text{k}}_l(\zeta_2b)
-\zeta_2 \, \bar{\text{k}}_l(\zeta_3b) \text{k}_l(\zeta_2b)}
{ \zeta_3 \, \text{i}_l(\zeta_3b) \bar{\text{k}}_l(\zeta_2b)
-\zeta_2 \, \bar{\text{i}}_l(\zeta_3b) \text{k}_l(\zeta_2b)},
\hspace{5mm}
\end{eqnarray}%
\label{refcoe-sp-def}%
\end{subequations}%
in terms of the shorthand notation
\begin{equation}
\zeta_i = \frac{n}{a_0} \sqrt{\varepsilon_i\left(i\,n\frac{c}{a_0}\right)},
\quad i=1,3,2.
\label{Imfreq-def}
\end{equation}
The temperature dependent constant $a_0$
that appears in Eq.\,(\ref{Imfreq-def})
%and Eq.\,(\ref{enSpL}) 
was introduced in Eq.\,(\ref{natscd}).
The reflection coefficients are expressed in terms of the modified 
spherical Bessel functions $\text{i}_l(t)$ and $\text{k}_l(t)$
that are related to the modified Bessel functions by the relations
\begin{subequations}
\begin{align}
\text{i}_l(t) &= \sqrt{\frac{\pi}{2t}} I_{l+\frac{1}{2}}(t), \\
\text{k}_l(t) &= \sqrt{\frac{\pi}{2t}} K_{l+\frac{1}{2}}(t).
\end{align}%
\label{msbf-def}%
\end{subequations}%
In particular $\text{i}_l(t) = \text{i}^{(1)}_l(t)$, 
the modified spherical Bessel function of the first kind, 
together with $\text{k}_l(t)$ are a suitable
pair of solutions in the right half of the complex plane
\cite{NIST:DLMF,NIST:2010fm}.
The respective functions with a bar are the generalized derivatives
of the modified spherical Bessel functions given by
\begin{subequations}
\begin{align}
\bar{\text{i}}_l(t) 
&= \bigg( \frac{1}{t} + \frac{\partial}{\partial t} \bigg) \text{i}_l(t), \\
\bar{\text{k}}_l(t) 
&= \bigg( \frac{1}{t} + \frac{\partial}{\partial t} \bigg) \text{k}_l(t).
\end{align}%
\label{bar-mBf}%
\end{subequations}%
Using the Wronskian for the modified spherical Bessel functions,
\begin{equation}
\text{k}_l \text{i}_l^\prime -\text{i}_l \text{k}_l^\prime = \frac{\pi}{2t^2},
\end{equation}
where primes denote differentiation, we have the relation 
\begin{equation}
\text{k}_l \bar{\text{i}}_l -\text{i}_l \bar{\text{k}}_l = \frac{\pi}{2t^2}.
\end{equation}
The reflection coefficients are frequently expressed in terms of the
modified Riccati-Bessel functions,
\begin{subequations}
\begin{align}
s_l(t) &= t \,\text{i}_l(t), \\ e_l(t) &= \frac{2t}{\pi}\,\text{k}_l(t),
\end{align} 
\end{subequations}
whose derivatives can be expressed in the form 
\begin{subequations}
\begin{align}
s_l^\prime(t) &= t \,\bar{\text{i}}_l(t), \\
e_l^\prime(t) &= \frac{2t}{\pi} \,\bar{\text{k}}_l(t).
\end{align}
\end{subequations}
For completeness we have provided the derivation of the Lifshitz energy
for concentric spherical configurations, given in Eq.\,(\ref{enSpL}),
in the Appendix.%\ref{sec-inten-spheres}

%-------------------------------------------------
\section{Asymptotic expansions}

Consider the scenario in which we know the dielectric functions
in Eq.\,(\ref{diel-depD}),
for the three media in Fig.~\ref{fig-con-spheres-12},
as a function of Matsubara mode number $n$ to a reasonable accuracy.
The computation of the interaction energy in Eq.\,(\ref{enSpL})
then, in principle, involves the evaluation of the sums
over the Matsubara mode number $n$ and the angular momentum mode number $l$. 
Both these sums contribute negligibly for large values of $n$ and $l$.
However, the reflection coefficients in Eqs.\,(\ref{refcoe-sp-def})
involve ratios of differences, and these differences get exceedingly
small for larger values of $l$. Thus, 
one has to keep an excessive number of significant digits in the
evaluation of the Bessel functions, which is computationally expensive.
This difficulty is avoided by expressing the modified
Bessel functions using (uniform) asymptotic expansions for large
order~\cite{NIST:DLMF,NIST:2010fm}.

%----------------
\begin{figure}
\includegraphics[width=7cm]{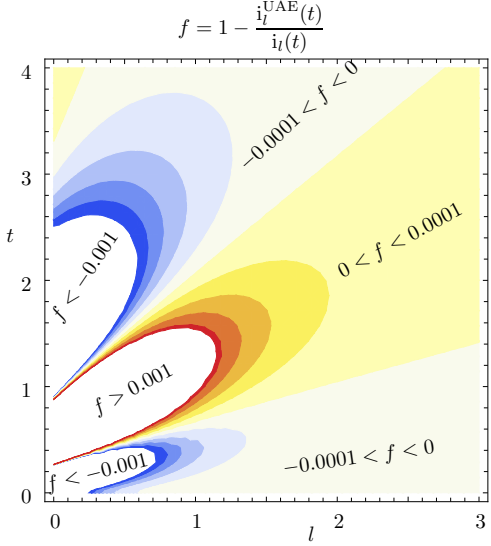}
\caption{Contour plot of fractional error $f$ in estimating
modified spherical Bessel function $\text{i}_l(t)$
using uniform asymptotic expansions. The errors are less
than one percent for $l<2$ and $t<3$, however, these errors
add up when we sum many terms.}
\label{fig-error-in-il}%
\end{figure}%
%----------------

The uniform asymptotic expansions for the modified spherical
Bessel functions are written using the definitions
\begin{equation}
\nu = l+\frac{1}{2}, \quad z = \frac{t}{\nu}, \quad
p(z) =\frac{1}{\sqrt{1+z^2}},
\end{equation}
and
\begin{equation}
\eta(z) = \sqrt{1+z^2} + \ln\left(\frac{z}{1+\sqrt{1+z^2}}\right),
\end{equation}
such that
\begin{subequations}
\begin{align}
\text{i}_l(t) &\sim \sqrt{\frac{p}{z}} 
\,\frac{e^{\nu\eta(z)}}{2\nu} \,A_\nu(p), \\
\text{k}_l(t) &\sim \pi \sqrt{\frac{p}{z}}
\,\frac{e^{-\nu\eta(z)}}{2\nu} \,B_\nu(p), \\
\bar{\text{i}}_l(t) &\sim \sqrt{\frac{1}{pz^3}} 
\,\frac{e^{\nu\eta(z)}}{2\nu} \,C_\nu(p), \\
\bar{\text{k}}_l(t) &\sim -\pi \sqrt{\frac{1}{pz^3}}
\,\frac{e^{-\nu\eta(z)}}{2\nu} \,D_\nu(p),
\end{align}%
\label{uae-msbf}%
\end{subequations}%
where
\begin{subequations}
\begin{align}
A_\nu(p) &\sim\sum_{k=0}^\infty \frac{u_k(p)}{\nu^k}, \\
B_\nu(p) &\sim\sum_{k=0}^\infty (-1)^k\frac{u_k(p)}{\nu^k}, \\
C_\nu(p) &\sim \sum_{k=0}^\infty \frac{v_k(p)}{\nu^k}
+\frac{p}{2\nu} \sum_{k=0}^\infty \frac{u_k(p)}{\nu^k}, \\
D_\nu(p) &\sim \sum_{k=0}^\infty (-1)^k\frac{v_k(p)}{\nu^k}
-\frac{p}{2\nu} \sum_{k=0}^\infty (-1)^k\frac{u_k(p)}{\nu^k},
\end{align}
\end{subequations}
are expressed in terms of polynomials generated by
\begin{subequations}
\begin{eqnarray}
u_{k+1}(p) &=& \frac{p^2(1-p^2)}{2} u_k^\prime(p)
+\int_0^p dq \frac{(1-5q^2)}{8} u_k(q), \\
v_{k+1}(p) &=& u_{k+1}(p) + p(p^2-1)
\left[ \frac{u_k(p)}{2} + p u_k^\prime(p) \right],  \hspace{10mm}
\end{eqnarray}
\end{subequations}
with $u_0(p)=1$ and $v_0(p)=1$.
The use of $\sim$ in place of equal sign in the equations
suggest that these involve asymptotic series and
the sums do not converge. 
The fractional error associated with
using the uniform asymptotic expansions for the modified 
spherical Bessel functions in Eqs.\,(\ref{uae-msbf})
is plotted in Fig.~\ref{fig-error-in-il} for order $l$ and
argument $t$. The fractional errors are small and the largest
error is only a percent for $l<2$ and $t<3$. Nevertheless, these
errors could add up to significant levels in the computation of energy.
This accumulation of error can be avoided in some cases by
keeping more terms in inverse powers of $\nu$ in the
sum on $k$, which is again computationally expensive.

Using the uniform asymptotic expansions for the modified spherical
Bessel functions in Eqs.\,(\ref{uae-msbf}) we derive the corresponding
expansions for the reflection coefficients in Eqs.\,(\ref{refcoe-sp-def})
to be
\begin{widetext}
\begin{subequations}
\begin{eqnarray}
r_{31}^E(a) &\sim& -\dfrac{1}{\pi} e^{2\nu\eta(\zeta_3a/\nu)}
\dfrac{p_1 A_\nu(p_1) C_\nu(p_3) -p_3 C_\nu(p_1) A_\nu(p_3)}
 {p_1 A_\nu(p_1) D_\nu(p_3) +p_3 C_\nu(p_1) B_\nu(p_3)}
\Bigg|_{p_i=p(\zeta_ia/\nu)}, \\
r_{31}^H(a) &\sim& -\dfrac{1}{\pi} e^{2\nu\eta(\zeta_3a/\nu)}
\dfrac{\zeta_1^2p_1 A_\nu(p_1) C_\nu(p_3) -\zeta_3^2p_3 C_\nu(p_1) A_\nu(p_3)}
 {\zeta_1^2p_1 A_\nu(p_1) D_\nu(p_3) +\zeta_3^2p_3 C_\nu(p_1) B_\nu(p_3)}
\Bigg|_{p_i=p(\zeta_ia/\nu)}, \\
r_{32}^E(b) &\sim& \pi e^{-2\nu\eta(\zeta_3b/\nu)}
\dfrac{p_3 B_\nu(p_3) D_\nu(p_2) -p_2 D_\nu(p_3) B_\nu(p_2)}
 {p_3 A_\nu(p_3) D_\nu(p_2) +p_2 C_\nu(p_3) B_\nu(p_2)}
\Bigg|_{p_i=p(\zeta_ib/\nu)}, \\
r_{32}^H(b) &\sim& \pi e^{-2\nu\eta(\zeta_3b/\nu)}
\dfrac{\zeta_3^2p_3 B_\nu(p_3) D_\nu(p_2) -\zeta_2^2p_2 D_\nu(p_3) B_\nu(p_2)}
 {\zeta_3^2p_3 A_\nu(p_3) D_\nu(p_2) +\zeta_2^2p_2 C_\nu(p_3) B_\nu(p_2)}
\Bigg|_{p_i=p(\zeta_ib/\nu)}.
\end{eqnarray}%
\label{uae-refcoe}%
\end{subequations}%
%\end{widetext}
The zero Matsubara mode, $n=0$, requires special consideration,
and is evaluated to be
\begin{subequations}
\begin{eqnarray}
r_{31}^E(a) r_{32}^E(b) \Big|_{n=0} &=& 0, \\
r_{31}^H(a) r_{32}^H(b) \Big|_{n=0} &=& 
-l(l+1) \left( \frac{a}{b} \right)^{2l+1}
%\nonumber \\ &&
\frac{[\varepsilon_1(0)-\varepsilon_3(0)]} 
 {[l\varepsilon_1(0)+(l+1)\varepsilon_3(0)]}
\frac{[\varepsilon_3(0)-\varepsilon_2(0)]} 
 {[l\varepsilon_2(0)+(l+1)\varepsilon_2(0)]}. 
\end{eqnarray}%
\label{refcoe-mzero}%
\end{subequations}%
\end{widetext}
Using these uniform asymptotic expansions for large order for the reflection
coefficients for non-zero Matsubara modes in Eqs.\,(\ref{uae-refcoe})
and the explicit evaluation of the zero Matsubara mode 
in Eq.\,(\ref{refcoe-mzero}),
in the expression for Lifshitz energy in Eq.\,(\ref{enSpL}),
we successfully circumvent the difficulty posed with numerically
evaluating quantities that involve very small numbers.

%-------------------------------------------------
\section{Numerical Procedure}

%----------------
%\begin{widetext}
\begin{table*}[t]
\begin{tabular}{| p{20mm}| p{20mm}| p{20mm}| p{20mm}| }
\hline
\rput[r](-0.5,-0.1){$1\,a_0$} \rput[r](-0.5,-0.4){\tiny (1.33\,$\mu$m)} 
\rput(1,0){$+1.99 \times 10^{-10}$} \rput(1,-0.4){\tiny (2,40,<1s,3)} 
& \rput(1,0){$+1.95 \times 10^{-7}$} \rput(1,-0.4){\tiny (2,40,<1s,3)} 
& \rput(1,0){$+1.55 \times 10^{-4}$} \rput(1,-0.4){\tiny (9,40,<1s,3)} 
& \rput(1,0){$+0.0416$} \rput(1,-0.4){\tiny (60,40,2s,3)} 
\\[5.5mm] \hline
\rput[br](-2,0){\psline[linewidth=0.5pt]{->}(0,0.5)} 
\rput[br](-1.93,0.6){$d$} 
\rput[r](-0.5,-0.1){$0.1\,a_0$} \rput[r](-0.5,-0.4){\tiny (133\,nm)} 
\rput(1,0){$-4.89\times 10^{-7}$} \rput(1,-0.4){\tiny (2,40,<1s,3)} 
& \rput(1,0){$-3.40 \times 10^{-4}$} \rput(1,-0.4){\tiny (2,40,<1s,3)} 
& \rput(1,0){$-0.0532$} \rput(1,-0.4){\tiny (9,40,<1s,3)} 
& \rput(1,0){$-3.68$} \rput(1,-0.4){\tiny (60,40,2s,3)} 
\\[5.5mm] \hline
\rput[r](-0.5,-0.1){$0.01\,a_0$} \rput[r](-0.5,-0.4){\tiny (13.3\,nm)} 
\rput(1,0){$-1.32 \times 10^{-3}$} \rput(1,-0.4){\tiny (2,350,1s,3)} 
& \rput(1,0){$-0.401$} \rput(1,-0.4){\tiny (6,350,2s,3)} 
& \rput(1,0){$-34.8$} \rput(1,-0.4){\tiny (45,350,12s,3)} 
& \rput(1,0){$-3.19 \times 10^3$} \rput(1,-0.4){\tiny (350,350,2m,3)} 
\\[5.5mm] \hline
\rput[r](-0.5,-0.1){$0.005\,a_0$} \rput[r](-0.5,-0.4){\tiny (6.65\,nm)}
\rput(1,0){$-4.19 \times 10^{-3}$} \rput(1,-0.4){\tiny (2,700,1s,3)} 
& \rput(1,0){$-1.18$} \rput(1,-0.4){\tiny (10,700,6s,3)} 
& \rput(1,0){$-107$} \rput(1,-0.4){\tiny (100,700,1m,3)} 
& \rput(1,0){$-1.02 \times 10^4$} \rput(1,-0.4){\tiny (700,700,7m,3)} 
\\[5.5mm] \hline
\rput[r](-0.5,-0.1){$0.001\,a_0$} \rput[r](-0.5,-0.4){\tiny (1.33\,nm)} 
\rput(1,0){$+0.143$} \rput(1,-0.4){\tiny (6,1400,7s,3)} 
\rput[t](1,-1.0){$0.001\,a_0$} \rput[t](1,-1.3){\tiny (1.33\,nm)} 
& \rput(1,0){$+9.1$} \rput(1,-0.4){\tiny (50,1500,1m,2)} 
\rput[t](1,-1.0){$0.01\,a_0$} \rput[t](1,-1.3){\tiny (13.3\,nm)} 
& \rput(1,0){$+8.2 \times 10^2$} \rput(1,-0.4){\tiny (400,1500,10m,2)} 
\rput[t](1,-1.0){$0.1\,a_0$} \rput[t](1,-1.3){\tiny (133\,nm)} 
& \rput(1,0){$+8.2 \times 10^4$} \rput(1,-0.4){\tiny (4000,2000,150m,2)} 
\rput[t](1,-1.0){$1\,a_0$} \rput[t](1,-1.3){\tiny (1.33\,$\mu$m)} 
\rput[tl](2,-1.1){\psline[linewidth=0.5pt]{->}(0.5,0)} 
\rput[tl](2.6,-1.03){$a$} 
\\[5.5mm] \hline 
\end{tabular}
\vspace{12mm}
\caption{
Numerical data for the
Lifshitz interaction energy $E(a,b)$, in Eq.\,(\ref{enSpL}),
in units of $E_0=k_BT/(2\pi)=\hbar c/(4\pi^2 a_0)$,
 for three concentric dielectric regions,
demarcated by radii $a$ and $b=a+d$, are catalogued with the respective
$l_\text{max}$ and $n_\text{max}$ to obtain convergence and confidence
in the necessary significant digits. Here $a_0=\hbar c/(2\pi k_BT)$,
which at $T=273.16$\,K yields $a_0\approx 1.3342\,\mu$m
and $E_0=k_BT/(2\pi) 
\approx 6.0023 \times 10^{-22}\,\text{J} 
\approx 3.7463\,\text{meV}$.
The numbers displayed in the tiny font in each box denote 
($l_\text{max}$, $n_\text{max}$, time, significant digits), for
ice-water-vapor configuration.
The maximum values for $l$ and $n$ are the values needed for the energy values
to converge to the required significant digits.
The time displayed is that for a typical personal computer. }
\label{table-nmaxlmax}
\end{table*}
%\end{widetext}
%----------------

The series in $l$ and $m$ are slowly converging even after
employing uniform asymptotic expansions.
We do not use any existing algorithms to speed up this slow
convergence. We simply sum the terms. Nevertheless, we report
the procedure in detail here for the sake of reproducibility.
The numerical work presented here is not state of the art,
and can be improved.

Our primary purpose in this article is to demonstrate that the Lifshitz
energy for concentric spherical configurations can be computed easily.
To this end, as an illustrative example, we consider a configuration
consisting of solid ice inside liquid water inside water vapor
in the configuration of Fig.~\ref{fig-con-spheres-12}.

To compute the numerical value for the Lifshitz energy in Eq.\,(\ref{enSpL})
we use the uniform asymptotic expansions for the
reflection coefficients given in Eqs.\,(\ref{uae-refcoe}),
instead of the exact expressions for the reflection coefficients
in Eqs.\,(\ref{refcoe-sp-def}).
This involves a sum on the Matsubara mode $n$, a sum on the angular
momentum mode $l$, and multiple sums on $k$ to generate the energy.
All these sums run from 0 to infinity,
but only an optimal number of terms are to be included to avoid
the unavoidable divergence associated with these asymptotic series.
Further, numerical computation can not sum infinite terms,
and dropping terms after an upper limit
in the sums introduces only an acceptable error. 
In this article we shall obtain convergence and confidence
in the numerical estimate up to three significant digits.

%----------------
\begin{figure}
\includegraphics{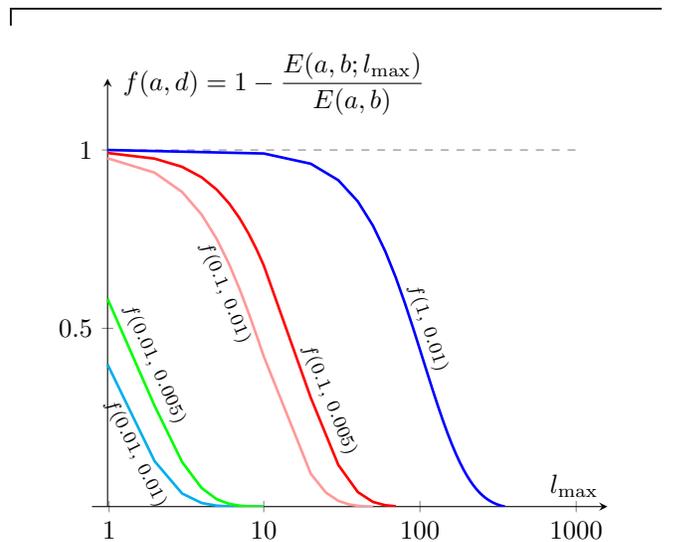}
\caption{Convergence of energy $E(a,b)$ to a fixed value,
during its computation. This is illustrated by plotting
the fraction $f(a,d)$ as a function
of the number of terms $l_\text{max}$ in the sum on $l$.
Recall $b=a+d$.
The numbers in the labels for the plot are in units of $a_0$. }
\label{fig-UAE-Energy-convergence}%
\end{figure}%
%----------------

We shall globally at the outset set
an upper limit in sums on $k$ of inverse powers of $\nu=l+1/2$
for the asymptotic expansions to be $k_\text{max}=3$.
It will be convenient to similarly set a global upper limit for the sum
on the Matsubara mode number, $n_\text{max}$, and for the sum on the
angular momentum mode, $l_\text{max}$. However, we learned that
the required upper limit varies widely for the different combinations
of the radii $a$ and $b$. For example, for the case
when the inner radius of the sphere is large and the difference in the outer
and inner radii is small, the sums on $n$ and $l$ in
the interaction energy of Eq.\,(\ref{enSpL}) need to be evaluated 
at least until $n_\text{max}=2000$ and $l_\text{max}=6000$
to obtain convergence and confidence in the data up to
three significant digits. Computationally this amounts to adding
$n_\text{max} \times l_\text{max} = 12 \times 10^6$ terms
in Eq.\,(\ref{enSpL}).
A typical personal computer takes about one millisecond to evaluate
one term in Eq.\,(\ref{enSpL}). 
To be specific, we used a computer with 
processor Intel Core i7-4700MQ CPU @ 2.40\,GHz\,$\times$\,8,
memory of 7.6\,GB, which amounts to about 100~GFLOPS,
and used Wolfram Mathematica~\cite{Mathematica} for evaluation.
Mathematica was preferred over other programs because
of the convenience to invoke libraries of special functions.
Thus, it takes a total of three hours
to evaluate the energy for one particular configuration.
The estimate for this time reduces by half
if the demand in accuracy is brought down to two significant digits.
To study the dependence of the energy in the two radii one
needs to at least compute the energy on a $10\times 10$ array
in the two radii. This amounts to three hundred hours of computation.
Though it is not impractical to proceed ahead, such long computation hours
makes the analysis tedious and inconvenient.
Nevertheless, the difficulty in calculating the energy for a particular
configuration is not as arduous as portrayed above.
One makes the observation that the computational burden is considerably
lower because the values of $n_\text{max}$ and $l_\text{max}$
needed for the necessary accuracy are significantly smaller for spherical
configurations of smaller radii.

Our strategy was to catalogue the $n_\text{max}$ and $l_\text{max}$
for all possible combinations of the radii.
This involves multiple runs to verify the convergence. However,
once catalogued it helps the analysis tremendously, because only
a small sector in the array is expensive on computational power.
The catalogue for the ice-water-vapor configuration
has been prepared in Table~\ref{table-nmaxlmax}.
We observe that the time taken to evaluate the energy for a particular
configuration is most often negligible. It is only when the inner
radii is large and difference in the radii is small that the time
is painstakingly long. The energies in Table~\ref{table-nmaxlmax}
are reported in units of $E_0=k_BT/(2\pi)=\hbar c/(4\pi^2 a_0)$
which is about 3.7463\,meV at the triple point of water, $T=273.16$\,K.
We illustrate the convergence of the energy for a particular values
of $a$ and $d$ as a function of the choice in $l_\text{max}$
in Fig.~\ref{fig-UAE-Energy-convergence}.
The convergence in energy
is computationally expensive in the bottom right corner of the
chart in Table~\ref{table-nmaxlmax}.

We made checks on the energies evaluated 
using uniform asymptotic expansions for the modified
spherical Bessel functions by comparing it with values for energy
obtained using the Bessel functions defined in Mathematica.
Remarkably,
to within three significant digits, the two results are identical
for the parameter space used in this study. We verified this extensively
for most of the parameter space, except for the few cases
with large radii of ice and small thickness of water for which
case the uniform asymptotic expansions fares very well.

It should be emphasized that the $l_\text{max}$ and $n_\text{max}$
presented in Table~\ref{table-nmaxlmax} is specific to the 
ice-water-vapor geometry. We expect the specific numbers
to be different for another set of dielectric materials. However, we 
expect the pattern to be similar. That is, for all materials larger
inner radii and small difference in radii will require the most
computational effort.

%-------------------------------------------------
\section{Elbaum-Schick effect in spherical geometry}
\label{sec-ES-sphere}

%----------------
\begin{figure}
\includegraphics[width=9.5cm]{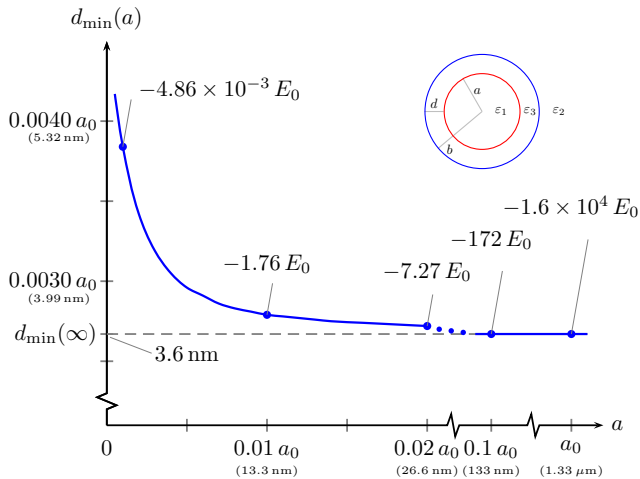}
\caption{For a spherical shell of water, immersed in vapor
and enclosing a ball of ice of radius $a$, at the triple point
of water, we plot the thickness $d_\text{min}(a)$ of the
shell of water that minimizes the Lifshitz energy for fixed $a$.
The equilibrium thickness of the water layer, $d_\text{min}(a)$,
is the least for large ice balls and increases for small radii
of ice balls. The equilibrium thickness of water
for the planar configuration $d_\text{min}(\infty)$ is achieved
to within $1\,\%$ for ice ball of radius 20\,nm.
The corresponding Lifshitz energy at equilibrium is marked
on the plot. The Lifshitz energy is minimum for large radii
of ice. For reference, $a_0\approx 1.33\,\mu\text{m}$ at the
triple point of water and $E_0=k_BT/(2\pi)=\hbar c/(4\pi^2 a_0)
\approx 6.0023\times 10^{-22}\,\text{J} 
\approx 3.7463\,\text{meV}$. }
\label{fig-mind-versus-a}%
\end{figure}%
%----------------

In Sec.~\ref{sec-EB-planar} we summarized how
Elbaum and Schick in Ref.~\cite{Elbaum1991wi} showed that at the triple
point of water, at equilibrium, it is energetically favorable 
to form a $3.6$\,nm thick layer of liquid water at a 
planar interface of solid ice and water vapor.
We shall use $d_\text{min}(\infty)$ to denote this thickness.
This is a delicate effect 
due to the fine differences in the frequency dependent
polarizabilities of ice and water and their interplay
in the presence of quantum fluctuations.
It is also a relativistic
effect in the sense that the effect is washed out if the analysis
does not accommodate retardation.
We inquire if the formation of a thin layer of water at
the interface of ice and vapor will be disturbed if the 
ice-vapor interface were curved. 
In Sec.~\ref{sec-inc-sur-mel} we pointed out that
difficulties in the experimental
verification of incomplete surface melting in the
experiment by Elbaum et al.~\cite{Elbaum:1993ims}
could be due to an unevenly flat interface.
Thus, the curvature dependence of the incomplete surface
melting is desired. 

We investigate
if it is energetically favorable to form a layer of water on
the surface of solid ice in the shape of a sphere of radius $a$ when it is 
immersed in water vapor of infinite extent at the triple point of water.
We find that a ball of solid ice at the triple point of water, at equilibrium,
permits a thicker layer of water to be formed on its surface,
relative to perfectly flat surface analyzed by Elbaum and Schick.
In Fig.~\ref{fig-mind-versus-a} we plot the thickness of liquid
water at equilibrium $d_\text{min}(a)$ as a function of
the radius of the ice ball $a$.
We observe that a ball of ice of 20\,nm radius is large enough
that in this context we can assume its surface to be sufficiently
flat for it to permit a water layer of thickness
$d_\text{min}(\infty)$ to within 1\% accuracy, with
the strength of instability decided by the binding energy of
about $-5E_0$. Here $E_0=k_BT/(2\pi)$ is a measure of the
quantum of energy available in the heat bath surrounding the system.

%----------------
\begin{figure}
\includegraphics[width=8cm]{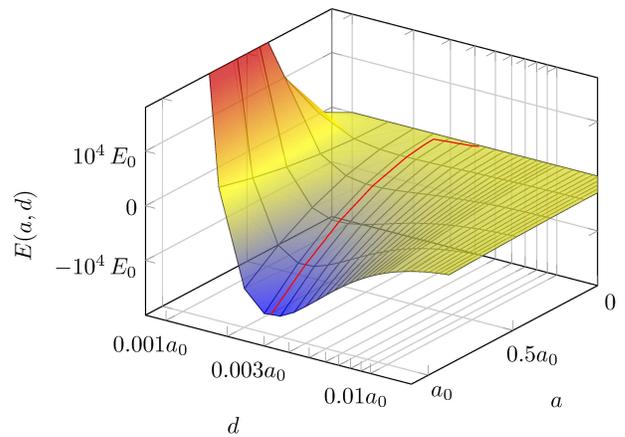}
\caption{
Lifshitz interaction energy $E(a,b)$, in Eq.\,(\ref{enSpL}),
in units of $E_0=k_BT/(2\pi)=\hbar c/(4\pi^2 a_0)$,
 for three concentric dielectric regions,
demarcated by radii $a$ and $b=a+d$,
such that inner medium represents solid ice, the intermediate medium
is liquid water, and the outer medium is water vapor,
plotted as a function of $a$ and $d=b-a$.
Here $a_0=\hbar c/(2\pi k_BT)$ is about $1.3342\,\mu\text{m}$ at the triple
point of water.
The red curve on the energy surface represents the thickness of water
layer at equilibrium, $d_\text{min}(a)$, plotted in
Fig.~\ref{fig-mind-versus-a} separately.
}
\label{fig-en-versus-ad}
\end{figure}%
%----------------

In Fig.~\ref{fig-mind-versus-a} the smallest radius of the ball
of ice we consider is 1.33\,nm ($0.001\,a_0$). In the range of
radii we have studied the thickness of water layer formed 
at equilibrium
monotonously increases for smaller radii of ice. Extrapolating this
behavior to zero radii of ice we conclude that a water layer of
infinite thickness is favored for small radii.
The divergence associated with $a\to 0$ is very weak and
consistent with our heuristic estimate
$d_\text{min} =-\ln(\bar\omega_{c2}a)/{\bar\omega_{c2}}$
for $a\ll d$, where 
$\bar\omega_{c2} =2\omega_{c2}\sqrt{\varepsilon_3(i\omega_{c2})}$,
similar to the estimate for the planar case in
Eq.\,(\ref{dmin-esP}).
In Fig.~\ref{fig-mind-versus-a} we also mark the energies associated
with each configuration. 
We observe that for radii of ice less than 10\,nm the energy associated
with the strength of instability is less than the 
quantum of energy available in the 
surrounding heat bath, which means that  for these cases the
disturbances in the surroundings will disturb the system and
the conclusions of Fig.~\ref{fig-mind-versus-a} are not relevant.
To gain a better insight of the
preference in energy we plot the Lifshitz interaction energy $E(a,b)$ 
as a function of both the radii of ice $a$ and the thickness of
the water layer $d$ as a three dimensional plot
in Fig.~\ref{fig-en-versus-ad}.
We also overlap the curve representing $d_\text{min}(a)$
in Fig.~\ref{fig-mind-versus-a} as a red curve on the energy surface
in Fig.~\ref{fig-en-versus-ad} for visual assistance.
For most part this curve remains constant in $d$ in the scale
of Fig.~\ref{fig-en-versus-ad} and starts diverging
for small radii of ice.
Slices in the three dimensional plot of Fig.~\ref{fig-en-versus-ad}
representing fixed $a$ are energy plots whose
minima are $d_\text{min}(a)$.

%-----------------------------------------------
\subsection{Promotion of ice formation in water}

In Fig.~\ref{fig-en-versus-ad} it is clear that the configuration
of minimum energy is for large radius of ice with a water layer having
a thickness $d_\text{min}(\infty)$. We also conclude that a spherical
drop of water inside an infinite extent of vapor
with no ice inside the water has zero interaction energy,
which can be concluded by extrapolating
the energies on the curve in Fig.~\ref{fig-mind-versus-a}.
This verifies that the Lifshitz interaction energy of
Eq.\,(\ref{enSpL}) is zero for $a=0$, or for $d\to\infty$. 
Thus, a drop of water surrounded by vapor at the triple point of water
is not stable. If ice is nucleated, the effects we consider will
help promote the growth in size of the ice region indefinitely with
the water layer thickness approaching $d_\text{min}(\infty)$.

An extrapolation leads to the hypothesis that zero point
energy could induce nucleation of ice in water. 
This is a remarkable proposition, because the common wisdom is that
ice nucleation requires an impurity like dust or soot or bacteria.
The suggestion is that quantum fluctuations
could contribute to inducing
nucleation of ice even in the absence of impurities.
However, the inward directed tension force on a small sphere
is strong, and considerable energy is required in order to make
nucleation possible. For a recent article on nucleation,
one may consult Ref.\,\cite{Espinosa:2019lnh}
and further references therein.

%-----------------------------------------------
\subsection{Superheating and supercooling}

Superheating of solids is the suspension of melting above the
melting point. Stranski in 1942~\cite{Stranski:1942sh} argued that
since superheating of solids is rarely observed
the surface of solids must be wetted by its liquid phase.
This argument is consistent with the idea of surface melting.

Supercooling of liquids is the absence of freezing
below the melting point. In striking contrast supercooling of
liquids is very common. It is well known that supercooled water
can exist as small droplets in clouds. 
This seems to be consistent with the conclusion that 
zero point energy alone is insufficient to induce nucleation
of ice in a water drop. However, 
the promotion of ice growth by the Lifshitz effect is more
pronounced for a big drop of water because of the relatively
large binding energy in Fig.~\ref{fig-en-versus-ad},
while for small drops of water
of 10\,nm and below the binding energy is too low and
less than the quantum of energy available in the surrounding heat bath.
Thus, it seems it should be easier to supercool small droplets
of water and harder to supercool big drops of water,
which is consistent with the observations.

%-----------------------------------------------
\subsection{Proximity force approximation}

%----------------
\begin{figure}
\includegraphics[width=8.0cm]{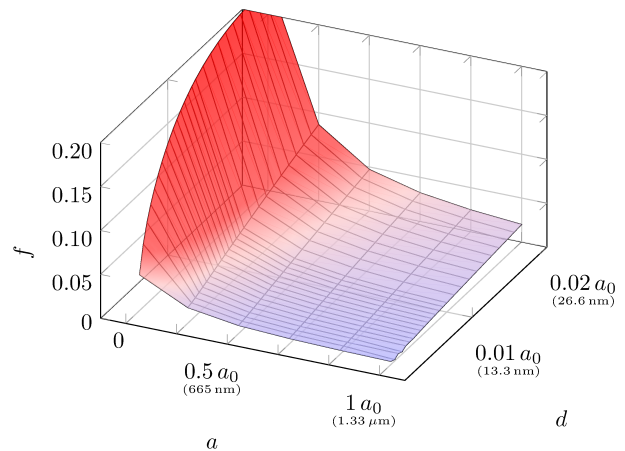}
\caption{Fractional difference in the Lifshitz energy $E(a,b)$
per surface area $4\pi a^2$ of the inner sphere of radius $a$
of a shell of water of thickness $d$
engulfing an ice ball of radius $a$,
with respect to the Lifshitz energy per unit area ${\cal E}$ for
a planar geometry, plotted as a function of $a$ and $d$.
The fractional error
$f = 1 - \frac{E(a,b)}{4\pi a^2} \frac{1}{{\cal E}}$
is plotted. The plots verify the general understanding that
the Lifshitz energy per unit area for a spherical
configuration approaches that of a planar configuration for
small thickness $d$. }
\label{fig-frac-error}
\end{figure}%
%----------------

It is often convenient to approximate the Lifshitz energy for
the configuration of concentric spheres with the corresponding
Lifshitz energy for planar configuration scaled with a suitable
area. This is often called the proximity force approximation,
and is usually a good approximation when the thickness of the 
intermediate medium is small compared to the radii of the
inner and outer spheres. 
In Fig.~\ref{fig-frac-error} we plot the fractional error
in using this approximation. This error is small for large
radii of ice and small thickness of water, and the error
gets significant for small radii of ice and large thickness
of water layer.

%-------------------------------------------------
\section{Conclusion and outlook}

In this article, we
successfully demonstrated that the Lifshitz energy for 
concentric spherical configurations in Fig.~\ref{fig-con-spheres-12}
can be computed with relative ease. As an application we
considered the case of solid ice enclosed by liquid water
inside water vapor at the triple point of water, and thereby
extended the analysis of Elbaum and Schick in Ref.~\cite{Elbaum1991wi}
to spherical interfaces. Our study shows that a drop of water
surrounded by vapor, with no ice inside the water, is unstable,
and quantum fluctuations promote formation of ice in the drop of
water at the triple point of water. It is energetically favorable for
the ice to grow indefinitely inside the drop
of water while a 3.6\,nm thick layer of water encircles the ball
of ice.
These conclusions ignore self-energies of the interior and
exterior spherical regions, which are not uniquely defined.
Some of these effects may be subsumed into surface tension,
but this omission, unavoidable at this stage of our understanding,
renders our conclusions tentative.
As noted, the effects we are considering are relatively small
compared to nucleation and surface tension effects.

In a following paper we will investigate the configuration of
water inside ice inside vapor. That is, is it energetically
favorable for ice to form at the interface of water and vapor,
and once formed will it grow inwards? 
This is of interest because in Ref.~\cite{Elbaum:1991sf}
it was found that no ice is formed on a planar water surface
based on Lifshitz theory.
This is expected to hold for water drops of large radii. 
In addition, now, we have in the present work found,
surprisingly, that purely quantum fluctuations promote freezing
from within water droplets instead of freezing from outside.

As an application of the results found here, that ice grows inside
water at the triple point of water, we would like to investigate the
relevance of this effect to the predictions for liquid water
on distant planets and their moons.
In presence of a silica surface we have predicted
that ice can form in water based on
Lifshitz theory\,\cite{Bostrom:2017isw}.
Bostr\"om et al. further proposed that Lifshitz forces could
lead to ice formation on some specific gas hydrate
surfaces in water~\cite{Bostrom:2019wwp}.
On some hypothesized ice coated oceans on the moons
Enceladus and Europa such ice films growing on CO$_2$
gas hydrate clusters could, if present, induce a size
dependent buoyancy for nanosized hydrate
clusters~\cite{Bostrom:2019wwp}.

Understanding the charging process of
atmospheric ice particles~\cite{SPWettlaufer2006}
is expected to be a relevant application of the results 
here.\footnote{This is a subject that one of the authors
(KAM) studied as a high-school student in 1961.}
Another application will involve studying ice formation in pores,
especially inside rocks and plants,
in light of our results here. 
In some situations double layer interaction energy be significant
in comparison to the Lifshitz energy
as was discussed in Ref.~\cite{Thiyam2018},
which could be extended to the spherical geometry.

%%%%%%-------------------------------------------------
%%%%%%-------------------------------------------------
\acknowledgments

We thank Clas Persson, Priyadarshini Thiyam, Oleksandr Malyi,
Kristian Berland, Johannes Fiedler, Stefan Buhmann, and Duston Wetzel,
for collaborations
on projects closely related to the discussions reported here.
We acknowledge support from The Research Council of Norway
(Project No. 250346) and United States National Science Foundation
(Grant No. 1707511).

%-----------------------------------------------------
%\bibliographystyle{plainnat}
%\bibliographystyle{unsrtnat}
\bibliography{biblio/b20160616-elbaum-schick}

\begin{thebibliography}{45}%
\makeatletter
\providecommand \@ifxundefined [1]{%
 \@ifx{#1\undefined}
}%
\providecommand \@ifnum [1]{%
 \ifnum #1\expandafter \@firstoftwo
 \else \expandafter \@secondoftwo
 \fi
}%
\providecommand \@ifx [1]{%
 \ifx #1\expandafter \@firstoftwo
 \else \expandafter \@secondoftwo
 \fi
}%
\providecommand \natexlab [1]{#1}%
\providecommand \enquote  [1]{``#1''}%
\providecommand \bibnamefont  [1]{#1}%
\providecommand \bibfnamefont [1]{#1}%
\providecommand \citenamefont [1]{#1}%
\providecommand \href@noop [0]{\@secondoftwo}%
\providecommand \href [0]{\begingroup \@sanitize@url \@href}%
\providecommand \@href[1]{\@@startlink{#1}\@@href}%
\providecommand \@@href[1]{\endgroup#1\@@endlink}%
\providecommand \@sanitize@url [0]{\catcode `\\12\catcode `\$12\catcode
  `\&12\catcode `\#12\catcode `\^12\catcode `\_12\catcode `\%12\relax}%
\providecommand \@@startlink[1]{}%
\providecommand \@@endlink[0]{}%
\providecommand \url  [0]{\begingroup\@sanitize@url \@url }%
\providecommand \@url [1]{\endgroup\@href {#1}{\urlprefix }}%
\providecommand \urlprefix  [0]{URL }%
\providecommand \Eprint [0]{\href }%
\providecommand \doibase [0]{http://dx.doi.org/}%
\providecommand \selectlanguage [0]{\@gobble}%
\providecommand \bibinfo  [0]{\@secondoftwo}%
\providecommand \bibfield  [0]{\@secondoftwo}%
\providecommand \translation [1]{[#1]}%
\providecommand \BibitemOpen [0]{}%
\providecommand \bibitemStop [0]{}%
\providecommand \bibitemNoStop [0]{.\EOS\space}%
\providecommand \EOS [0]{\spacefactor3000\relax}%
\providecommand \BibitemShut  [1]{\csname bibitem#1\endcsname}%
\let\auto@bib@innerbib\@empty
%</preamble>
\bibitem [{\citenamefont {van~der Waals}(1873)}]{Waals:1873sl}%
  \BibitemOpen
  \bibfield  {author} {\bibinfo {author} {\bibfnamefont {J.~D.}\ \bibnamefont
  {van~der Waals}},\ }\emph {\bibinfo {title} {{Over de Continuiteit van den
  Gas-en Vloeistoftoestand, ({O}n the continuity of the gas and liquid
  state)}}},\ \href@noop {} {Ph.D. thesis},\ \bibinfo  {school} {Universiteit
  Leiden (Leiden University)}, \bibinfo {address} {{The Netherlands}} (\bibinfo
  {year} {1873})\BibitemShut {NoStop}%
\bibitem [{\citenamefont {Eisenschitz}\ and\ \citenamefont
  {London}(1930)}]{London:1930a}%
  \BibitemOpen
  \bibfield  {author} {\bibinfo {author} {\bibfnamefont {R.}~\bibnamefont
  {Eisenschitz}}\ and\ \bibinfo {author} {\bibfnamefont {F.}~\bibnamefont
  {London}},\ }\bibfield  {title} {\enquote {\bibinfo {title} {{\"{U}}ber das
  {V}rh{\"{a}}ltnis der van der {W}aalsschen {K}r{\"{a}}fte zu den
  {H}om{\"{o}}opolaren {B}indungskr{\"{a}}ften},}\ }\href@noop {} {\bibfield
  {journal} {\bibinfo  {journal} {Z. Physik}\ }\textbf {\bibinfo {volume}
  {60}},\ \bibinfo {pages} {491} (\bibinfo {year} {1930})},\ \bibinfo {note}
  {english translation in \cite{Hettema:2001cq}}\BibitemShut {NoStop}%
\bibitem [{\citenamefont {London}(1930)}]{London:1930b}%
  \BibitemOpen
  \bibfield  {author} {\bibinfo {author} {\bibfnamefont {F.}~\bibnamefont
  {London}},\ }\bibfield  {title} {\enquote {\bibinfo {title} {Zur {T}heorie
  und {S}ystematik der {M}olekularkr{\"{a}}fte},}\ }\href@noop {} {\bibfield
  {journal} {\bibinfo  {journal} {Z. Physik}\ }\textbf {\bibinfo {volume}
  {63}},\ \bibinfo {pages} {245} (\bibinfo {year} {1930})},\ \bibinfo {note}
  {english translation in \cite{Hettema:2001cq}}\BibitemShut {NoStop}%
\bibitem [{\citenamefont {Hettema}(2000)}]{Hettema:2001cq}%
  \BibitemOpen
  \bibfield  {author} {\bibinfo {author} {\bibfnamefont {H.}~\bibnamefont
  {Hettema}},\ }\href@noop {} {\emph {\bibinfo {title} {Quantum Chemistry:
  {C}lassic Scientific Papers}}},\ World Scientific Series in 20th century
  chemistry\ (\bibinfo  {publisher} {World Scientific},\ \bibinfo {year}
  {2000})\BibitemShut {NoStop}%
\bibitem [{\citenamefont {Casimir}\ and\ \citenamefont
  {Polder}(1948)}]{Casmir:1947hx}%
  \BibitemOpen
  \bibfield  {author} {\bibinfo {author} {\bibfnamefont {H.~B.~G.}\
  \bibnamefont {Casimir}}\ and\ \bibinfo {author} {\bibfnamefont
  {D.}~\bibnamefont {Polder}},\ }\bibfield  {title} {\enquote {\bibinfo {title}
  {{The influence of retardation on the London-van der Waals forces}},}\ }\href
  {\doibase 10.1103/PhysRev.73.360} {\bibfield  {journal} {\bibinfo  {journal}
  {Phys. Rev.}\ }\textbf {\bibinfo {volume} {73}},\ \bibinfo {pages} {360}
  (\bibinfo {year} {1948})}\BibitemShut {NoStop}%
%%CITATION = PHRVA,73,360;%%
\bibitem [{\citenamefont {Planck}(1914)}]{Planck:1914rht}%
  \BibitemOpen
  \bibfield  {author} {\bibinfo {author} {\bibfnamefont {M.}~\bibnamefont
  {Planck}},\ }\href@noop {} {\emph {\bibinfo {title} {The Theory of Heat
  Radiation}}}\ (\bibinfo  {publisher} {P. Blakiston's Son \& Co.},\ \bibinfo
  {address} {Philadelphia},\ \bibinfo {year} {1914})\ \bibinfo {note}
  {authorized translation by M. Masius}\BibitemShut {NoStop}%
\bibitem [{\citenamefont {Einstein}\ and\ \citenamefont
  {Hopf}(1910)}]{Einstein:1910ser}%
  \BibitemOpen
  \bibfield  {author} {\bibinfo {author} {\bibfnamefont {A.}~\bibnamefont
  {Einstein}}\ and\ \bibinfo {author} {\bibfnamefont {L.}~\bibnamefont
  {Hopf}},\ }\bibfield  {title} {\enquote {\bibinfo {title}
  {\href{https://doi.org/10.1002/andp.19103381604} {Statistische Untersuchung
  der Bewegung eines Resonators in einem Strahlungsfeld}},}\ }\href@noop {}
  {\bibfield  {journal} {\bibinfo  {journal} {Ann. Phys.}\ }\textbf {\bibinfo
  {volume} {338}},\ \bibinfo {pages} {1105} (\bibinfo {year}
  {1910})}\BibitemShut {NoStop}%
\bibitem [{\citenamefont {Einstein}\ and\ \citenamefont
  {Stern}(1913)}]{Einstein:1913nab}%
  \BibitemOpen
  \bibfield  {author} {\bibinfo {author} {\bibfnamefont {A.}~\bibnamefont
  {Einstein}}\ and\ \bibinfo {author} {\bibfnamefont {O.}~\bibnamefont
  {Stern}},\ }\bibfield  {title} {\enquote {\bibinfo {title}
  {\href{https://doi.org/10.1002/andp.19133450309} {Einige Argumente für die
  Annahme einer molekularen Agitation beim absoluten Nullpunkt}},}\ }\href@noop
  {} {\bibfield  {journal} {\bibinfo  {journal} {Ann. Phys.}\ }\textbf
  {\bibinfo {volume} {345}},\ \bibinfo {pages} {551} (\bibinfo {year}
  {1913})}\BibitemShut {NoStop}%
\bibitem [{\citenamefont {Planck}(1901)}]{Planck:1901sne}%
  \BibitemOpen
  \bibfield  {author} {\bibinfo {author} {\bibfnamefont {M.}~\bibnamefont
  {Planck}},\ }\bibfield  {title} {\enquote {\bibinfo {title}
  {\href{https://doi.org/10.1002/andp.19013090310} {Ueber das Gesetz der
  Energieverteilung im Normalspectrum}},}\ }\href@noop {} {\bibfield  {journal}
  {\bibinfo  {journal} {Ann. Phys.}\ }\textbf {\bibinfo {volume} {309}},\
  \bibinfo {pages} {553} (\bibinfo {year} {1901})}\BibitemShut {NoStop}%
\bibitem [{\citenamefont {Casimir}(1948)}]{Casimir:1948pc}%
  \BibitemOpen
  \bibfield  {author} {\bibinfo {author} {\bibfnamefont {H.~B.~G.}\
  \bibnamefont {Casimir}},\ }\bibfield  {title} {\enquote {\bibinfo {title}
  {{On the attraction between two perfectly conducting plates}},}\ }\href@noop
  {} {\bibfield  {journal} {\bibinfo  {journal} {Kon. Ned. Akad. Wetensch.
  Proc.}\ }\textbf {\bibinfo {volume} {51}},\ \bibinfo {pages} {793} (\bibinfo
  {year} {1948})}\BibitemShut {NoStop}%
\bibitem [{\citenamefont {Lifshitz}(1956)}]{Lifshitz:1956sb}%
  \BibitemOpen
  \bibfield  {author} {\bibinfo {author} {\bibfnamefont {E.~M.}\ \bibnamefont
  {Lifshitz}},\ }\bibfield  {title} {\enquote {\bibinfo {title}
  {\href{http://www.jetp.ac.ru/cgi-bin/e/index/e/2/1/p73?a=list} {The theory of
  molecular attractive forces between solids}},}\ }\href@noop {} {\bibfield
  {journal} {\bibinfo  {journal} {Sov. Phys. JETP}\ }\textbf {\bibinfo {volume}
  {2}},\ \bibinfo {pages} {73} (\bibinfo {year} {1956})},\ \bibinfo {note}
  {[Translated from: Zh. Eksp. Teor. Fiz. {\bf 29}, 94 (1956)]}\BibitemShut
  {NoStop}%
\bibitem [{\citenamefont {Dzyaloshinskii}\ \emph {et~al.}(1961)\citenamefont
  {Dzyaloshinskii}, \citenamefont {Lifshitz},\ and\ \citenamefont
  {Pitaevskii}}]{Dzyaloshinskii:1961fw}%
  \BibitemOpen
  \bibfield  {author} {\bibinfo {author} {\bibfnamefont {I.~E.}\ \bibnamefont
  {Dzyaloshinskii}}, \bibinfo {author} {\bibfnamefont {E.~M.}\ \bibnamefont
  {Lifshitz}}, \ and\ \bibinfo {author} {\bibfnamefont {L.~P.}\ \bibnamefont
  {Pitaevskii}},\ }\bibfield  {title} {\enquote {\bibinfo {title}
  {\href{http://stacks.iop.org/0038-5670/4/i=2/a=R01} {General theory of van
  der {W}aals' forces}},}\ }\href@noop {} {\bibfield  {journal} {\bibinfo
  {journal} {Soviet Physics Uspekhi}\ }\textbf {\bibinfo {volume} {4}},\
  \bibinfo {pages} {153} (\bibinfo {year} {1961})},\ \bibinfo {note}
  {[Translated from: Usp. Fiz. Nauk {\bf 73}, 381 (1961)]}\BibitemShut
  {NoStop}%
\bibitem [{\citenamefont {Elbaum}\ and\ \citenamefont
  {Schick}(1991{\natexlab{a}})}]{Elbaum1991wi}%
  \BibitemOpen
  \bibfield  {author} {\bibinfo {author} {\bibfnamefont {M.}~\bibnamefont
  {Elbaum}}\ and\ \bibinfo {author} {\bibfnamefont {M.}~\bibnamefont
  {Schick}},\ }\bibfield  {title} {\enquote {\bibinfo {title}
  {\href{http://doi.org/10.1103/PhysRevLett.66.1713} {Application of the theory
  of dispersion forces to the surface melting of ice}},}\ }\href@noop {}
  {\bibfield  {journal} {\bibinfo  {journal} {Phys. Rev. Lett.}\ }\textbf
  {\bibinfo {volume} {66}},\ \bibinfo {pages} {1713} (\bibinfo {year}
  {1991}{\natexlab{a}})}\BibitemShut {NoStop}%
\bibitem [{\citenamefont {Thiyam}\ \emph
  {et~al.}(2018{\natexlab{a}})\citenamefont {Thiyam}, \citenamefont {Parashar},
  \citenamefont {Shajesh}, \citenamefont {Malyi}, \citenamefont {Bostr\"om},
  \citenamefont {Milton}, \citenamefont {Brevik},\ and\ \citenamefont
  {Persson}}]{Thiyam:2018lsr}%
  \BibitemOpen
  \bibfield  {author} {\bibinfo {author} {\bibfnamefont {P.}~\bibnamefont
  {Thiyam}}, \bibinfo {author} {\bibfnamefont {P.}~\bibnamefont {Parashar}},
  \bibinfo {author} {\bibfnamefont {K.~V.}\ \bibnamefont {Shajesh}}, \bibinfo
  {author} {\bibfnamefont {O.~I.}\ \bibnamefont {Malyi}}, \bibinfo {author}
  {\bibfnamefont {M.}~\bibnamefont {Bostr\"om}}, \bibinfo {author}
  {\bibfnamefont {K.~A.}\ \bibnamefont {Milton}}, \bibinfo {author}
  {\bibfnamefont {I.}~\bibnamefont {Brevik}}, \ and\ \bibinfo {author}
  {\bibfnamefont {C.}~\bibnamefont {Persson}},\ }\bibfield  {title} {\enquote
  {\bibinfo {title} {\href{https://doi.org/10.1103/PhysRevLett.120.131601}
  {Distance-dependent sign-reversal in the Casimir-Lifshitz torque}},}\
  }\href@noop {} {\bibfield  {journal} {\bibinfo  {journal} {Phys. Rev. Lett.}\
  }\textbf {\bibinfo {volume} {120}},\ \bibinfo {pages} {131601} (\bibinfo
  {year} {2018}{\natexlab{a}})}\BibitemShut {NoStop}%
\bibitem [{\citenamefont {Kenneth}\ and\ \citenamefont
  {Klich}(2006)}]{Kenneth:2006vr}%
  \BibitemOpen
  \bibfield  {author} {\bibinfo {author} {\bibfnamefont {O.}~\bibnamefont
  {Kenneth}}\ and\ \bibinfo {author} {\bibfnamefont {I.}~\bibnamefont
  {Klich}},\ }\bibfield  {title} {\enquote {\bibinfo {title}
  {\href{https://doi.org/10.1103/PhysRevLett.97.160401} {Opposites attract: A
  Theorem about the {C}asimir force}},}\ }\href@noop {} {\bibfield  {journal}
  {\bibinfo  {journal} {Phys. Rev. Lett.}\ }\textbf {\bibinfo {volume} {97}},\
  \bibinfo {pages} {160401} (\bibinfo {year} {2006})}\BibitemShut {NoStop}%
\bibitem [{\citenamefont {Shajesh}\ and\ \citenamefont
  {Schaden}(2011)}]{Shajesh:2011ef}%
  \BibitemOpen
  \bibfield  {author} {\bibinfo {author} {\bibfnamefont {K.~V.}\ \bibnamefont
  {Shajesh}}\ and\ \bibinfo {author} {\bibfnamefont {M.}~\bibnamefont
  {Schaden}},\ }\bibfield  {title} {\enquote {\bibinfo {title}
  {\href{http://doi.org/10.1103/PhysRevD.83.125032} {Many-body contributions to
  Green's functions and Casimir energies}},}\ }\href@noop {} {\bibfield
  {journal} {\bibinfo  {journal} {Phys. Rev. D}\ }\textbf {\bibinfo {volume}
  {83}},\ \bibinfo {pages} {125032} (\bibinfo {year} {2011})}\BibitemShut
  {NoStop}%
\bibitem [{\citenamefont {Krech}(1994)}]{Krech:1994}%
  \BibitemOpen
  \bibfield  {author} {\bibinfo {author} {\bibfnamefont {M.}~\bibnamefont
  {Krech}},\ }\href@noop {} {\emph {\bibinfo {title} {Casimir Effect in
  Critical Systems}}}\ (\bibinfo  {publisher} {World Scientific},\ \bibinfo
  {address} {Singapore},\ \bibinfo {year} {1994})\BibitemShut {NoStop}%
\bibitem [{\citenamefont {Fiedler}\ \emph {et~al.}(2019)\citenamefont
  {Fiedler}, \citenamefont {Bostr{\"o}m}, \citenamefont {Persson},
  \citenamefont {Brevik}, \citenamefont {Corkery}, \citenamefont {Buhmann},\
  and\ \citenamefont {Parsons}}]{Fiedler:2019ddw}%
  \BibitemOpen
  \bibfield  {author} {\bibinfo {author} {\bibfnamefont {J.}~\bibnamefont
  {Fiedler}}, \bibinfo {author} {\bibfnamefont {M.}~\bibnamefont
  {Bostr{\"o}m}}, \bibinfo {author} {\bibfnamefont {C.}~\bibnamefont
  {Persson}}, \bibinfo {author} {\bibfnamefont {I.}~\bibnamefont {Brevik}},
  \bibinfo {author} {\bibfnamefont {R.~W.}\ \bibnamefont {Corkery}}, \bibinfo
  {author} {\bibfnamefont {S.~Y.}\ \bibnamefont {Buhmann}}, \ and\ \bibinfo
  {author} {\bibfnamefont {D.~F.}\ \bibnamefont {Parsons}},\ }\href@noop {} {\
  (\bibinfo {year} {2019})},\ \bibinfo {note} {under preparation}\BibitemShut
  {NoStop}%
\bibitem [{\citenamefont {Bostr\"om}\ \emph {et~al.}(2017)\citenamefont
  {Bostr\"om}, \citenamefont {Malyi}, \citenamefont {Parashar}, \citenamefont
  {Shajesh}, \citenamefont {Thiyam}, \citenamefont {Milton}, \citenamefont
  {Persson}, \citenamefont {Parsons},\ and\ \citenamefont
  {Brevik}}]{Bostrom:2017isw}%
  \BibitemOpen
  \bibfield  {author} {\bibinfo {author} {\bibfnamefont {M.}~\bibnamefont
  {Bostr\"om}}, \bibinfo {author} {\bibfnamefont {O.~I.}\ \bibnamefont
  {Malyi}}, \bibinfo {author} {\bibfnamefont {P.}~\bibnamefont {Parashar}},
  \bibinfo {author} {\bibfnamefont {K.~V.}\ \bibnamefont {Shajesh}}, \bibinfo
  {author} {\bibfnamefont {P.}~\bibnamefont {Thiyam}}, \bibinfo {author}
  {\bibfnamefont {K.~A.}\ \bibnamefont {Milton}}, \bibinfo {author}
  {\bibfnamefont {C.}~\bibnamefont {Persson}}, \bibinfo {author} {\bibfnamefont
  {D.~F.}\ \bibnamefont {Parsons}}, \ and\ \bibinfo {author} {\bibfnamefont
  {I.}~\bibnamefont {Brevik}},\ }\bibfield  {title} {\enquote {\bibinfo {title}
  {\href{https://doi.org/10.1103/PhysRevB.95.155422} {Lifshitz interaction can
  promote ice growth at water-silica interfaces}},}\ }\href@noop {} {\bibfield
  {journal} {\bibinfo  {journal} {Phys. Rev. B}\ }\textbf {\bibinfo {volume}
  {95}},\ \bibinfo {pages} {155422} (\bibinfo {year} {2017})}\BibitemShut
  {NoStop}%
\bibitem [{\citenamefont {Weyl}(1951)}]{Weyl:1951is}%
  \BibitemOpen
  \bibfield  {author} {\bibinfo {author} {\bibfnamefont {W.~A.}\ \bibnamefont
  {Weyl}},\ }\bibfield  {title} {\enquote {\bibinfo {title}
  {\href{https://doi.org/10.1016/0095-8522(51)90011-6} {Surface structure of
  water and some of its physical and chemical manifestations}},}\ }\href@noop
  {} {\bibfield  {journal} {\bibinfo  {journal} {J. Colloid Sci.}\ }\textbf
  {\bibinfo {volume} {6}},\ \bibinfo {pages} {389} (\bibinfo {year}
  {1951})}\BibitemShut {NoStop}%
\bibitem [{\citenamefont {Fletcher}(1968)}]{Fletcher:1968is}%
  \BibitemOpen
  \bibfield  {author} {\bibinfo {author} {\bibfnamefont {N.~H.}\ \bibnamefont
  {Fletcher}},\ }\bibfield  {title} {\enquote {\bibinfo {title}
  {\href{https://doi.org/10.1080/14786436808227758} {Surface structure of water
  and ice}},}\ }\href@noop {} {\bibfield  {journal} {\bibinfo  {journal}
  {Philos. Mag. A}\ }\textbf {\bibinfo {volume} {18}},\ \bibinfo {pages} {1287}
  (\bibinfo {year} {1968})}\BibitemShut {NoStop}%
\bibitem [{\citenamefont {Dash}\ \emph {et~al.}(1995)\citenamefont {Dash},
  \citenamefont {Fu},\ and\ \citenamefont {Wettlaufer}}]{Dash:1995cei}%
  \BibitemOpen
  \bibfield  {author} {\bibinfo {author} {\bibfnamefont {J.~G.}\ \bibnamefont
  {Dash}}, \bibinfo {author} {\bibfnamefont {H.}~\bibnamefont {Fu}}, \ and\
  \bibinfo {author} {\bibfnamefont {J.~S.}\ \bibnamefont {Wettlaufer}},\
  }\bibfield  {title} {\enquote {\bibinfo {title}
  {\href{https://doi.org/10.1088/0034-4885/58/1/003} {The premelting of ice and
  its environmental consequences}},}\ }\href@noop {} {\bibfield  {journal}
  {\bibinfo  {journal} {Rep. Prog. Phys.}\ }\textbf {\bibinfo {volume} {58}},\
  \bibinfo {pages} {115} (\bibinfo {year} {1995})}\BibitemShut {NoStop}%
\bibitem [{\citenamefont {Elbaum}\ \emph {et~al.}(1995)\citenamefont {Elbaum},
  \citenamefont {Lipson},\ and\ \citenamefont {Wettlaufer}}]{Elbaum:1995wcp}%
  \BibitemOpen
  \bibfield  {author} {\bibinfo {author} {\bibfnamefont {M.}~\bibnamefont
  {Elbaum}}, \bibinfo {author} {\bibfnamefont {S.~G.}\ \bibnamefont {Lipson}},
  \ and\ \bibinfo {author} {\bibfnamefont {J.~S.}\ \bibnamefont {Wettlaufer}},\
  }\bibfield  {title} {\enquote {\bibinfo {title}
  {\href{https://doi.org/10.1209/0295-5075/29/6/005} {Evaporation Preempts
  Complete Wetting}},}\ }\href@noop {} {\bibfield  {journal} {\bibinfo
  {journal} {EPL}\ }\textbf {\bibinfo {volume} {29}},\ \bibinfo {pages} {457}
  (\bibinfo {year} {1995})}\BibitemShut {NoStop}%
\bibitem [{\citenamefont {Slater}\ and\ \citenamefont
  {Michaelides}(2019)}]{Slater:2019ms}%
  \BibitemOpen
  \bibfield  {author} {\bibinfo {author} {\bibfnamefont {B.}~\bibnamefont
  {Slater}}\ and\ \bibinfo {author} {\bibfnamefont {A.}~\bibnamefont
  {Michaelides}},\ }\bibfield  {title} {\enquote {\bibinfo {title}
  {\href{https://doi.org/10.1038/s41570-019-0080-8} {Surface premelting of
  water ice}},}\ }\href@noop {} {\bibfield  {journal} {\bibinfo  {journal}
  {Nature Rev. Chem.}\ }\textbf {\bibinfo {volume} {3}},\ \bibinfo {pages}
  {172} (\bibinfo {year} {2019})}\BibitemShut {NoStop}%
\bibitem [{\citenamefont {Elbaum}\ \emph {et~al.}(1993)\citenamefont {Elbaum},
  \citenamefont {Lipson},\ and\ \citenamefont {Dash}}]{Elbaum:1993ims}%
  \BibitemOpen
  \bibfield  {author} {\bibinfo {author} {\bibfnamefont {M.}~\bibnamefont
  {Elbaum}}, \bibinfo {author} {\bibfnamefont {S.~G.}\ \bibnamefont {Lipson}},
  \ and\ \bibinfo {author} {\bibfnamefont {J.~G.}\ \bibnamefont {Dash}},\
  }\bibfield  {title} {\enquote {\bibinfo {title}
  {\href{https://doi.org/10.1016/0022-0248(93)90483-D} {Optical study of
  surface melting on ice}},}\ }\href@noop {} {\bibfield  {journal} {\bibinfo
  {journal} {J. Cryst. Growth}\ }\textbf {\bibinfo {volume} {129}},\ \bibinfo
  {pages} {491} (\bibinfo {year} {1993})}\BibitemShut {NoStop}%
\bibitem [{\citenamefont {Casimir}(1953)}]{Casimir:1953eqr}%
  \BibitemOpen
  \bibfield  {author} {\bibinfo {author} {\bibfnamefont {H.~B.~G.}\
  \bibnamefont {Casimir}},\ }\bibfield  {title} {\enquote {\bibinfo {title}
  {\href{https://doi.org/10.1016/S0031-8914(53)80095-9} {Introductory remarks
  on quantum electrodynamics}},}\ }\href@noop {} {\bibfield  {journal}
  {\bibinfo  {journal} {Physica}\ }\textbf {\bibinfo {volume} {19}},\ \bibinfo
  {pages} {846} (\bibinfo {year} {1953})}\BibitemShut {NoStop}%
\bibitem [{\citenamefont {Boyer}(1968)}]{Boyer:1968uf}%
  \BibitemOpen
  \bibfield  {author} {\bibinfo {author} {\bibfnamefont {T.~H.}\ \bibnamefont
  {Boyer}},\ }\bibfield  {title} {\enquote {\bibinfo {title}
  {\href{https://doi.org/10.1103/PhysRev.174.1764} {Quantum electromagnetic
  zero point energy of a conducting spherical shell and the Casimir model for a
  charged particle}},}\ }\href@noop {} {\bibfield  {journal} {\bibinfo
  {journal} {Phys. Rev.}\ }\textbf {\bibinfo {volume} {174}},\ \bibinfo {pages}
  {1764} (\bibinfo {year} {1968})}\BibitemShut {NoStop}%
\bibitem [{\citenamefont {Milton}(1980)}]{Milton1980bc}%
  \BibitemOpen
  \bibfield  {author} {\bibinfo {author} {\bibfnamefont {K.~A.}\ \bibnamefont
  {Milton}},\ }\bibfield  {title} {\enquote {\bibinfo {title}
  {\href{https://doi.org/10.1016/0003-4916(80)90149-9} {Semiclassical electron
  models: Casimir self-stress in dielectric and conducting balls}},}\
  }\href@noop {} {\bibfield  {journal} {\bibinfo  {journal} {Ann. Phys. (N.
  Y.)}\ }\textbf {\bibinfo {volume} {127}},\ \bibinfo {pages} {49} (\bibinfo
  {year} {1980})}\BibitemShut {NoStop}%
\bibitem [{\citenamefont {Brevik}\ and\ \citenamefont
  {Kolbenstvedt}(1983)}]{Brevik:1983mcd}%
  \BibitemOpen
  \bibfield  {author} {\bibinfo {author} {\bibfnamefont {I.}~\bibnamefont
  {Brevik}}\ and\ \bibinfo {author} {\bibfnamefont {H.}~\bibnamefont
  {Kolbenstvedt}},\ }\bibfield  {title} {\enquote {\bibinfo {title}
  {\href{https://doi.org/10.1016/0003-4916(83)90196-3} {Electromagnetic Casimir
  densities in dielectric spherical media}},}\ }\href@noop {} {\bibfield
  {journal} {\bibinfo  {journal} {Ann. Phys.}\ }\textbf {\bibinfo {volume}
  {149}},\ \bibinfo {pages} {237} (\bibinfo {year} {1983})}\BibitemShut
  {NoStop}%
\bibitem [{\citenamefont {Candelas}(1982)}]{Candelas1982sc}%
  \BibitemOpen
  \bibfield  {author} {\bibinfo {author} {\bibfnamefont {P.}~\bibnamefont
  {Candelas}},\ }\bibfield  {title} {\enquote {\bibinfo {title}
  {\href{https://doi.org/10.1016/0003-4916(82)90029-X} {Vacuum energy in the
  presence of dielectric and conducting surfaces}},}\ }\href@noop {} {\bibfield
   {journal} {\bibinfo  {journal} {Ann. Phys. (N. Y.)}\ }\textbf {\bibinfo
  {volume} {143}},\ \bibinfo {pages} {241} (\bibinfo {year}
  {1982})}\BibitemShut {NoStop}%
\bibitem [{\citenamefont {Brevik}\ \emph
  {et~al.}(2002{\natexlab{a}})\citenamefont {Brevik}, \citenamefont {Aarseth},\
  and\ \citenamefont {H\o{}ye}}]{Brevik2001am}%
  \BibitemOpen
  \bibfield  {author} {\bibinfo {author} {\bibfnamefont {I.}~\bibnamefont
  {Brevik}}, \bibinfo {author} {\bibfnamefont {J.~B.}\ \bibnamefont {Aarseth}},
  \ and\ \bibinfo {author} {\bibfnamefont {J.~S.}\ \bibnamefont {H\o{}ye}},\
  }\bibfield  {title} {\enquote {\bibinfo {title}
  {\href{https://arxiv.org/abs/quant-ph/0111037} {Casimir problem in spherical
  dielectrics: A quantum statistical mechanical approach}},}\ }\href@noop {}
  {\bibfield  {journal} {\bibinfo  {journal} {Int. J. Mod. Phys. A}\ }\textbf
  {\bibinfo {volume} {17}} (\bibinfo {year} {2002}{\natexlab{a}})}\BibitemShut
  {NoStop}%
\bibitem [{\citenamefont {H\o{}ye}\ \emph {et~al.}(2001)\citenamefont
  {H\o{}ye}, \citenamefont {Brevik},\ and\ \citenamefont
  {Aarseth}}]{Brevik2001at}%
  \BibitemOpen
  \bibfield  {author} {\bibinfo {author} {\bibfnamefont {J.~S.}\ \bibnamefont
  {H\o{}ye}}, \bibinfo {author} {\bibfnamefont {I.}~\bibnamefont {Brevik}}, \
  and\ \bibinfo {author} {\bibfnamefont {J.~B.}\ \bibnamefont {Aarseth}},\
  }\bibfield  {title} {\enquote {\bibinfo {title}
  {\href{http://doi.org/10.1103/PhysRevE.63.051101} {Casimir problem of
  spherical dielectrics: Quantum statistical and field theoretical
  approaches}},}\ }\href@noop {} {\bibfield  {journal} {\bibinfo  {journal}
  {Phys. Rev. E}\ }\textbf {\bibinfo {volume} {63}},\ \bibinfo {pages} {051101}
  (\bibinfo {year} {2001})}\BibitemShut {NoStop}%
\bibitem [{\citenamefont {Brevik}\ \emph
  {et~al.}(2002{\natexlab{b}})\citenamefont {Brevik}, \citenamefont {Aarseth},\
  and\ \citenamefont {H\o{}ye}}]{Brevik2002pg}%
  \BibitemOpen
  \bibfield  {author} {\bibinfo {author} {\bibfnamefont {I.}~\bibnamefont
  {Brevik}}, \bibinfo {author} {\bibfnamefont {J.~B.}\ \bibnamefont {Aarseth}},
  \ and\ \bibinfo {author} {\bibfnamefont {J.~S.}\ \bibnamefont {H\o{}ye}},\
  }\bibfield  {title} {\enquote {\bibinfo {title}
  {\href{http://doi.org/10.1103/PhysRevE.66.026119} {Casimir problem of
  spherical dielectrics: Numerical evaluation for general permittivities}},}\
  }\href@noop {} {\bibfield  {journal} {\bibinfo  {journal} {Phys. Rev. E}\
  }\textbf {\bibinfo {volume} {66}},\ \bibinfo {pages} {026119} (\bibinfo
  {year} {2002}{\natexlab{b}})}\BibitemShut {NoStop}%
\bibitem [{\citenamefont {Parashar}\ \emph {et~al.}(2017)\citenamefont
  {Parashar}, \citenamefont {Milton}, \citenamefont {Shajesh},\ and\
  \citenamefont {Brevik}}]{Parashar:2017sgo}%
  \BibitemOpen
  \bibfield  {author} {\bibinfo {author} {\bibfnamefont {P.}~\bibnamefont
  {Parashar}}, \bibinfo {author} {\bibfnamefont {K.~A.}\ \bibnamefont
  {Milton}}, \bibinfo {author} {\bibfnamefont {K.~V.}\ \bibnamefont {Shajesh}},
  \ and\ \bibinfo {author} {\bibfnamefont {I.}~\bibnamefont {Brevik}},\
  }\bibfield  {title} {\enquote {\bibinfo {title}
  {\href{https://doi.org/10.1103/PhysRevD.96.085010} {Electromagnetic
  $\delta$-function sphere}},}\ }\href@noop {} {\bibfield  {journal} {\bibinfo
  {journal} {Phys. Rev. D}\ }\textbf {\bibinfo {volume} {96}},\ \bibinfo
  {pages} {085010} (\bibinfo {year} {2017})}\BibitemShut {NoStop}%
\bibitem [{\citenamefont {Shajesh}\ \emph {et~al.}(2017)\citenamefont
  {Shajesh}, \citenamefont {Parashar},\ and\ \citenamefont
  {Brevik}}]{Shajesh:2017ssa}%
  \BibitemOpen
  \bibfield  {author} {\bibinfo {author} {\bibfnamefont {K.~V.}\ \bibnamefont
  {Shajesh}}, \bibinfo {author} {\bibfnamefont {P.}~\bibnamefont {Parashar}}, \
  and\ \bibinfo {author} {\bibfnamefont {I.}~\bibnamefont {Brevik}},\
  }\bibfield  {title} {\enquote {\bibinfo {title}
  {\href{https://doi.org/10.1016/j.aop.2017.10.008} {Casimir–Polder energy
  for axially symmetric systems}},}\ }\href@noop {} {\bibfield  {journal}
  {\bibinfo  {journal} {Ann. Phys. (N.Y.)}\ }\textbf {\bibinfo {volume}
  {387}},\ \bibinfo {pages} {166} (\bibinfo {year} {2017})}\BibitemShut
  {NoStop}%
\bibitem [{{\relax DLMF}()}]{NIST:DLMF}%
  \BibitemOpen
  {\relax DLMF},\ \href@noop {} {\enquote {\bibinfo {title}
  {\href{http://dlmf.nist.gov/} {NIST Digital Library of Mathematical
  Functions}},}\ }\bibinfo {howpublished} {Release 1.0.8 of 2014-04-25},\
  \bibinfo {note} {online companion to \cite{NIST:2010fm}}\BibitemShut
  {NoStop}%
\bibitem [{NIS(2010)}]{NIST:2010fm}%
  \BibitemOpen
  \enquote {\bibinfo {title} {\href{http://dlmf.nist.gov/} {{NIST} handbook of
  mathematical functions}},}\ \ (\bibinfo  {publisher} {Cambridge University
  Press},\ \bibinfo {address} {New York},\ \bibinfo {year} {2010})\ \bibinfo
  {note} {edited by F. W. J. Olver and D. W. Lozier and R. F. Boisvert and C.
  W. Clark. Print companion to \cite{NIST:DLMF}}\BibitemShut {NoStop}%
\bibitem [{\citenamefont {Inc.}()}]{Mathematica}%
  \BibitemOpen
  \bibfield  {author} {\bibinfo {author} {\bibfnamefont {Wolfram~Research{,}}\
  \bibnamefont {Inc.}},\ }\href@noop {} {\enquote {\bibinfo {title}
  {Mathematica, {V}ersion 11.1},}\ }\bibinfo {note} {Champaign, IL,
  2019}\BibitemShut {NoStop}%
\bibitem [{\citenamefont {Espinosa}\ \emph {et~al.}(2019)\citenamefont
  {Espinosa}, \citenamefont {Diez}, \citenamefont {Vega}, \citenamefont
  {Valeriani}, \citenamefont {Ramirez},\ and\ \citenamefont
  {Sanz}}]{Espinosa:2019lnh}%
  \BibitemOpen
  \bibfield  {author} {\bibinfo {author} {\bibfnamefont {J.~R.}\ \bibnamefont
  {Espinosa}}, \bibinfo {author} {\bibfnamefont {A.~L.}\ \bibnamefont {Diez}},
  \bibinfo {author} {\bibfnamefont {C.}~\bibnamefont {Vega}}, \bibinfo {author}
  {\bibfnamefont {C.}~\bibnamefont {Valeriani}}, \bibinfo {author}
  {\bibfnamefont {J.}~\bibnamefont {Ramirez}}, \ and\ \bibinfo {author}
  {\bibfnamefont {E.}~\bibnamefont {Sanz}},\ }\bibfield  {title} {\enquote
  {\bibinfo {title} {\href{http://dx.doi.org/10.1039/C8CP07432A} {Ice Ih vs.
  ice III along the homogeneous nucleation line}},}\ }\href@noop {} {\bibfield
  {journal} {\bibinfo  {journal} {Phys. Chem. Chem. Phys.}\ }\textbf {\bibinfo
  {volume} {21}},\ \bibinfo {pages} {5655} (\bibinfo {year}
  {2019})}\BibitemShut {NoStop}%
\bibitem [{\citenamefont {Stranski}(1942)}]{Stranski:1942sh}%
  \BibitemOpen
  \bibfield  {author} {\bibinfo {author} {\bibfnamefont {I.~N.}\ \bibnamefont
  {Stranski}},\ }\bibfield  {title} {\enquote {\bibinfo {title}
  {{\href{https://doi.org/10.1007/BF01476465} {\"{U}ber den Schmelzvorgang bei
  nichtpolaren Kristallen}, (On the melting process of nonpolar crystals)}},}\
  }\href@noop {} {\bibfield  {journal} {\bibinfo  {journal}
  {Naturwissenschaften}\ }\textbf {\bibinfo {volume} {30}},\ \bibinfo {pages}
  {425} (\bibinfo {year} {1942})}\BibitemShut {NoStop}%
\bibitem [{\citenamefont {Elbaum}\ and\ \citenamefont
  {Schick}(1991{\natexlab{b}})}]{Elbaum:1991sf}%
  \BibitemOpen
  \bibfield  {author} {\bibinfo {author} {\bibfnamefont {M.}~\bibnamefont
  {Elbaum}}\ and\ \bibinfo {author} {\bibfnamefont {M.}~\bibnamefont
  {Schick}},\ }\bibfield  {title} {\enquote {\bibinfo {title}
  {\href{https://doi.org/10.1051/jp1:1991233} {On the failure of water to
  freeze from its surface}},}\ }\href@noop {} {\bibfield  {journal} {\bibinfo
  {journal} {J. Phys. I France}\ }\textbf {\bibinfo {volume} {1}},\ \bibinfo
  {pages} {1665} (\bibinfo {year} {1991}{\natexlab{b}})}\BibitemShut {NoStop}%
\bibitem [{\citenamefont {Bostr\"om}\ \emph {et~al.}(2019)\citenamefont
  {Bostr\"om}, \citenamefont {Corkery}, \citenamefont {Lima}, \citenamefont
  {Malyi}, \citenamefont {Buhmann}, \citenamefont {Persson}, \citenamefont
  {Brevik}, \citenamefont {Parsons},\ and\ \citenamefont
  {Fiedler}}]{Bostrom:2019wwp}%
  \BibitemOpen
  \bibfield  {author} {\bibinfo {author} {\bibfnamefont {M.}~\bibnamefont
  {Bostr\"om}}, \bibinfo {author} {\bibfnamefont {R.~W.}\ \bibnamefont
  {Corkery}}, \bibinfo {author} {\bibfnamefont {E.~R.~A.}\ \bibnamefont
  {Lima}}, \bibinfo {author} {\bibfnamefont {O.~I.}\ \bibnamefont {Malyi}},
  \bibinfo {author} {\bibfnamefont {S.~Y.}\ \bibnamefont {Buhmann}}, \bibinfo
  {author} {\bibfnamefont {C.}~\bibnamefont {Persson}}, \bibinfo {author}
  {\bibfnamefont {I.}~\bibnamefont {Brevik}}, \bibinfo {author} {\bibfnamefont
  {D.~F.}\ \bibnamefont {Parsons}}, \ and\ \bibinfo {author} {\bibfnamefont
  {J.}~\bibnamefont {Fiedler}},\ }\bibfield  {title} {\enquote {\bibinfo
  {title} {\href{https://doi.org/10.1021/acsearthspacechem.9b00019} {Dispersion
  Forces Stabilize Ice Coatings at Certain Gas Hydrate Interfaces That Prevent
  Water Wetting}},}\ }\href@noop {} {\bibfield  {journal} {\bibinfo  {journal}
  {ACS Earth Space Chem.}\ }\textbf {\bibinfo {volume} {3}},\ \bibinfo {pages}
  {1014} (\bibinfo {year} {2019})}\BibitemShut {NoStop}%
\bibitem [{\citenamefont {Sherwood}\ \emph {et~al.}(2006)\citenamefont
  {Sherwood}, \citenamefont {Phillips},\ and\ \citenamefont
  {Wettlaufer}}]{SPWettlaufer2006}%
  \BibitemOpen
  \bibfield  {author} {\bibinfo {author} {\bibfnamefont {S.~C.}\ \bibnamefont
  {Sherwood}}, \bibinfo {author} {\bibfnamefont {V.~T.~J.}\ \bibnamefont
  {Phillips}}, \ and\ \bibinfo {author} {\bibfnamefont {J.~S.}\ \bibnamefont
  {Wettlaufer}},\ }\bibfield  {title} {\enquote {\bibinfo {title}
  {\href{https://doi.org/10.1029/2005GL025242} {Small ice crystals and the
  climatology of lightning}},}\ }\href@noop {} {\bibfield  {journal} {\bibinfo
  {journal} {Geophys. Res. Lett}\ }\textbf {\bibinfo {volume} {33}},\ \bibinfo
  {pages} {L05804} (\bibinfo {year} {2006})}\BibitemShut {NoStop}%
\bibitem [{\citenamefont {Thiyam}\ \emph
  {et~al.}(2018{\natexlab{b}})\citenamefont {Thiyam}, \citenamefont {Fiedler},
  \citenamefont {Buhmann}, \citenamefont {Persson}, \citenamefont {Brevik},
  \citenamefont {Bostr{\"o}m},\ and\ \citenamefont {Parsons}}]{Thiyam2018}%
  \BibitemOpen
  \bibfield  {author} {\bibinfo {author} {\bibfnamefont {P.}~\bibnamefont
  {Thiyam}}, \bibinfo {author} {\bibfnamefont {J.}~\bibnamefont {Fiedler}},
  \bibinfo {author} {\bibfnamefont {S.~Y.}\ \bibnamefont {Buhmann}}, \bibinfo
  {author} {\bibfnamefont {C.}~\bibnamefont {Persson}}, \bibinfo {author}
  {\bibfnamefont {I.}~\bibnamefont {Brevik}}, \bibinfo {author} {\bibfnamefont
  {M.}~\bibnamefont {Bostr{\"o}m}}, \ and\ \bibinfo {author} {\bibfnamefont
  {D.~F.}\ \bibnamefont {Parsons}},\ }\bibfield  {title} {\enquote {\bibinfo
  {title} {\href{https://doi.org/10.1021/acs.jpcc.8b02351} {Ice particles sink
  below the water surface due to a balance of salt, van der Waals, and buoyancy
  forces}},}\ }\href@noop {} {\bibfield  {journal} {\bibinfo  {journal} {J.
  Phys. Chem. C}\ }\textbf {\bibinfo {volume} {122}},\ \bibinfo {pages} {15311}
  (\bibinfo {year} {2018}{\natexlab{b}})}\BibitemShut {NoStop}%
\bibitem [{\citenamefont {Prudnikov}\ \emph {et~al.}(1990)\citenamefont
  {Prudnikov}, \citenamefont {Brychkov},\ and\ \citenamefont
  {Marichev}}]{Prudnikov:1990}%
  \BibitemOpen
  \bibfield  {author} {\bibinfo {author} {\bibfnamefont {A.~P.}\ \bibnamefont
  {Prudnikov}}, \bibinfo {author} {\bibfnamefont {Yu.~A.}\ \bibnamefont
  {Brychkov}}, \ and\ \bibinfo {author} {\bibfnamefont {O.~I.}\ \bibnamefont
  {Marichev}},\ }\href@noop {} {\emph {\bibinfo {title} {Integrals and
  Series}}},\ Vol.~\bibinfo {volume} {2}\ (\bibinfo  {publisher} {Gordon and
  Breach},\ \bibinfo {address} {Amsterdam},\ \bibinfo {year}
  {1990})\BibitemShut {NoStop}%
\end{thebibliography}%
%\nocite{*} %%% Will print the complete bib data.
%-----------------------------------------------------

%-----------------------------------------------------
\appendix*
%-----------------------------------------------------
\section{Lifshitz interaction energy for concentric spheres}
\label{sec-inten-spheres}

In this appendix we use $\hbar=1$ and $c=1$ 
for typographic brevity. This can be undone
by replacing $\zeta \to \zeta/c$ and introducing $\hbar$ in
equations for energy.

In the multiple scattering formalism the Lifshitz interaction energy
for the configuration of concentric spheres in
Fig.~\ref{fig-con-spheres-12} is given by
\begin{equation}
E_{12}(a,b) = \frac{1}{2} \int_{-\infty}^\infty \frac{d\zeta}{2\pi}
\,\text{Tr} \,\ln \Big[ {\bf 1} 
- {\bm\Gamma}_a V_a \cdot {\bm\Gamma}_b V_b \Big],
\label{LEcsI}
\end{equation}
where
\begin{subequations}
\begin{eqnarray}
V_a &=& (\varepsilon_3-1) 
+ (\varepsilon_1-\varepsilon_3) \theta(a-r), \\
V_b &=& (\varepsilon_3-1) 
+ (\varepsilon_2-\varepsilon_3) \theta(r-b),
\end{eqnarray}
\end{subequations}
each, describe concentric spherical regions with a single interface,
obtained by letting $b\to\infty$ and $a\to 0$, respectively,
in Fig.~\ref{fig-con-spheres-12}.
The interaction energy of Eq.\,(\ref{LEcsI}) corresponds to the
fourth term in the decomposition of energy in Eq.\,(\ref{edecom})
for the system in Fig.~\ref{fig-con-spheres-12}, which is
finite by construction.
In Eq.\,(\ref{LEcsI}) we used symbolic notation,
\begin{equation}
{\bm\Gamma}_a V_a \cdot {\bm\Gamma}_b V_b
= \int d^3\bar r\,
{\bm\Gamma}_a({\bf r},\bar{\bf r}) V_a(\bar{\bf r}) 
\cdot {\bm\Gamma}_b(\bar{\bf r},{\bf r}^\prime) V_b({\bf r}^\prime). 
\label{lokersy}
\end{equation}
Thus, the argument of the logarithm in Eq.\,(\ref{LEcsI}) is a 
dyadic, or a matrix, with elements constituting integral kernels.
The trace in Eq.\,(\ref{LEcsI}) is over the matrix indices
and on the kernel coordinates ${\bf r}$ and ${\bf r}^\prime$.
The Green dyadics ${\bm\Gamma}_a({\bf r},{\bf r}^\prime)$
and ${\bm\Gamma}_b({\bf r},{\bf r}^\prime)$ can be
suitably expressed in the basis of spherical vector
eigenfunctions~\cite{Shajesh:2017ssa}
\begin{subequations}
\begin{align}
{\bf X}_{lm}^{(u)}(\theta,\phi)
&= \frac{1}{ik_\perp}{\bm\nabla}_\perp
Y_{lm}(\theta,\phi), \label{Uefsp} \\
{\bf X}_{lm}^{(v)}(\theta,\phi)
&= \frac{1}{ik_\perp}
\hat{\bf r} \times {\bm\nabla}_\perp
Y_{lm}(\theta,\phi), \label{Vefsp} \\
{\bf X}_{lm}^{(w)}(\theta,\phi)
&= \hat{\bf r} \,
Y_{lm}(\theta,\phi), \label{Wefsp}
\end{align}%
\label{vec-EF-sp}%
\end{subequations}
expressed in terms of spherical harmonics
$Y_{lm}(\theta,\phi)$, as
\begin{eqnarray}
{\bm\Gamma}_\alpha({\bf r},{\bf r}^\prime) = 
\hspace{60mm} \nonumber \\ \hspace{5mm}
\sum_{l=0/1}^\infty \sum_{m=-l}^l 
{\bf X}_{lm}^{(i)}(\theta,\phi) \gamma_{lm,\alpha}^{ij}(r,r^\prime) 
{\bf X}_{lm}^{(j)*}(\theta^\prime,\phi^\prime),
\hspace{8mm}
\label{G=XgX-sp}%
\end{eqnarray}%
$\alpha=a,b$,
where $0/1$ for the initial value of index $l$ means that
the sum over $l$ runs from 0 to $\infty$ for terms
involving ${\bf X}_{lm}^{(w)}$, but $l$ runs from 1 to $\infty$ for
terms involving ${\bf X}_{lm}^{(u)}$ and ${\bf X}_{lm}^{(v)}$.
The matrices $\gamma_{lm,\alpha}^{ij}(r,r^\prime)$
are the components of the Green dyadics
in the basis of spherical vector eigenfunctions
given by
\begin{eqnarray}
\gamma_{lm,\alpha}^{ij}(r,r^\prime) =
\nonumber \hspace{60mm} \\
 \left[ \begin{array}{c} 
\frac{D}{\varepsilon_\alpha(r)}
\frac{D^\prime}{\varepsilon_\alpha(r^\prime)}
g_{l,\alpha}^H(r,r^\prime)
\hspace{3mm} 0 \hspace{3mm}
\frac{D}{\varepsilon_\alpha(r)}
\frac{ik_\perp^\prime}{\varepsilon_\alpha(r^\prime)}
 g_{l,\alpha}^H(r,r^\prime) \\[2mm]
0 \hspace{10mm} -\zeta^2 g_{l,\alpha}^E(r,r^\prime) \hspace{10mm} 0 \\[2mm]
\frac{-ik_\perp}{\varepsilon_\alpha(r)} 
\frac{D^\prime}{\varepsilon_\alpha(r^\prime)}
g_{l,\alpha}^H(r,r^\prime) 
\hspace{3mm} 0 \hspace{3mm}
\frac{-ik_\perp}{\varepsilon_\alpha(r)}
\frac{ik_\perp^\prime}{\varepsilon_\alpha(r^\prime)}
 g_{l,\alpha}^H(r,r^\prime)
\end{array} \right],
\hspace{8mm}
\label{Gamma=gE-sp}
\end{eqnarray}
%\begin{widetext}
%\begin{align}
%\gamma_{lm,\alpha}^{ij}(r,r^\prime)
%&= \left[ \begin{array}{ccc} \frac{1}{\varepsilon_\alpha(r)}
%\left( \frac{1}{r} +\frac{\partial}{\partial r} \right)
%\frac{1}{\varepsilon_\alpha(r^\prime)}
%\left( \frac{1}{r^\prime} +\frac{\partial}{\partial r^\prime} \right)
% g_{l,\alpha}^H(r,r^\prime) & 0 & \frac{1}{\varepsilon_\alpha(r)}
%\left( \frac{1}{r} +\frac{\partial}{\partial r} \right)
%\frac{ik_\perp^\prime}{\varepsilon_\alpha(r^\prime)}
% g_{l,\alpha}^H(r,r^\prime) \\[2mm]
%0 & \omega^2 g_{l,\alpha}^E(r,r^\prime) & 0 \\[2mm]
%-\frac{ik_\perp}{\varepsilon_\alpha(r)} \frac{1}{\varepsilon_\alpha(r^\prime)}
%\left( \frac{1}{r^\prime} +\frac{\partial}{\partial r^\prime} \right)
%g_{l,\alpha}^H(r,r^\prime) & 0 & -\frac{ik_\perp}{\varepsilon_\alpha(r)}
%\frac{ik_\perp^\prime}{\varepsilon_\alpha(r^\prime)}
% g_{l,\alpha}^H(r,r^\prime)
%\end{array} \right]
%\label{Gamma=gE-sp}
%\end{align}
%\end{widetext}
where
\begin{equation}
\varepsilon_a(r) = \begin{cases}
\varepsilon_1, & r<a, \\ \varepsilon_3, & a<r, \end{cases}
\end{equation}
and
\begin{equation}
\varepsilon_b(r) = \begin{cases}
\varepsilon_3, & r<b, \\ \varepsilon_2, & b<r, \end{cases}
\end{equation}
with shorthand notations
\begin{equation}
k_\perp^2 = \frac{l(l+1)}{r^2},
\qquad
k_\perp^{\prime\, 2} = \frac{l(l+1)}{{r^\prime}^2},
\label{kperpp-def}
\end{equation}
and
\begin{equation}
D = \left( \frac{1}{r} +\frac{\partial}{\partial r} \right)
= \frac{1}{r} \frac{\partial}{\partial r} r
\end{equation}
and similarly for $D^\prime$ with primed coordinates.
We have omitted a term containing a $\delta$-function 
in Eq.\,(\ref{Gamma=gE-sp}), which does not contribute to 
interaction energies between disjoint bodies.
The transverse magnetic and transverse electric
spherical Green's functions in Eq.\,(\ref{Gamma=gE-sp})
satisfy the differential equations
\begin{subequations}
\begin{eqnarray}
\left[ -D \frac{1}{\varepsilon_\alpha(r)}
D + \frac{l(l+1)}{r^2\varepsilon_\alpha(r)}
+ \zeta^2 \right] g_{l,\alpha}^H(r,r^\prime)
&=& \frac{\delta(r-r^\prime)}{r^2}, \\
\left[ -D^2 +\frac{l(l+1)}{r^2} + \zeta^2 \varepsilon_\alpha(r)
\right] g_{l,\alpha}^E(r,r^\prime) 
&=& \frac{\delta(r-r^\prime)}{r^2},
\hspace{14mm}
\label{EM-GfunH-sp}%
\end{eqnarray}%
\label{EM-Gfun-sp}%
\end{subequations}
$\alpha=a,b$, and have solutions
\begin{widetext}
\begin{subequations}
\begin{eqnarray}
g_{l,a}^H(r^\prime, r) 
&=& -\frac{\varepsilon_1}{a} \frac{1}{(\zeta_1a)}
\frac{\zeta_3 \, \text{i}_l (\zeta_1r) \text{k}_l (\zeta_3r^\prime) }
{ \left[ \zeta_1 \, \text{i}_l (\zeta_1a) \bar{\text{k}}_l (\zeta_3a)
-\zeta_3 \, \bar{\text{i}}_l(\zeta_1a) \text{k}_l(\zeta_3a) \right]}, \\
g_{l,b}^H(r,r^\prime) 
&=& -\frac{\varepsilon_2}{b} \frac{1}{(\zeta_2b)}
\frac{\zeta_3 \, \text{i}_l(\zeta_3r) \text{k}_l(\zeta_2r^\prime)}
{ \left[ \zeta_3 \, \text{i}_l(\zeta_3b) \bar{\text{k}}_l(\zeta_2b)
-\zeta_2 \, \bar{\text{i}}_l(\zeta_3b) \text{k}_l(\zeta_2b) \right]}, \\
g_{l,a}^E(r^\prime, r) 
&=& -\frac{1}{a} \frac{1}{(\zeta_1a)}
\frac{\zeta_1 \, \text{i}_l(\zeta_1r) \text{k}_l(\zeta_3r^\prime)}
{ \left[ \zeta_3 \, \text{i}_l(\zeta_1a) \bar{\text{k}}_l(\zeta_3a)
-\zeta_1 \, \bar{\text{i}}_l(\zeta_1a) \text{k}_l(\zeta_3a) \right]}, \\
g_{l,b}^E(r,r^\prime) 
&=& -\frac{1}{b} \frac{1}{(\zeta_2b)}
\frac{\zeta_2 \, \text{i}_l(\zeta_3r) \text{k}_l(\zeta_2r^\prime)}
{ \left[ \zeta_2 \, \text{i}_l(\zeta_3b) \bar{\text{k}}_l(\zeta_2b)
-\zeta_3 \, \bar{\text{i}}_l(\zeta_3b) \text{k}_l(\zeta_2b) \right]},
\hspace{5mm}
\end{eqnarray}%
\label{gfsolrrp}%
\end{subequations}%
\end{widetext}
in terms of modified spherical Bessel functions of
Eqs.\,(\ref{msbf-def}) and generalized derivatives of 
of modified spherical Bessel functions in 
Eqs.\,(\ref{bar-mBf}) with
\begin{equation}
\zeta_i = \zeta \sqrt{\varepsilon_i}.
\end{equation}

To evaluate the Lifshitz interaction energy we begin by 
processing the dyadic in Eq.\,(\ref{lokersy}). 
We use the expressions for the Green dyadics in 
Eq.\,(\ref{G=XgX-sp}) and using the orthogonality relations
for the spherical vector eigenfunctions,
\begin{eqnarray}
\int_0^\pi \sin\theta d\theta \int_0^{2\pi} d\phi\,
{\bf X}_{lm}^{*(i)}(\theta,\phi)
{\bf X}_{l^\prime m^\prime}^{(j)}(\theta,\phi)
\nonumber \hspace{15mm} \\
=\delta_{ll^\prime} \delta_{mm^\prime} \delta_{ij},
\hspace{50mm}
\end{eqnarray}
for the angular part of coordinate $\bar{\bf r}$, we obtain
\begin{eqnarray}
{\bm\Gamma}_a V_a \cdot {\bm\Gamma}_b V_b
= \sum_{l=0}^\infty \sum_{m=-l}^l
{\bf X}_{lm}^{(i)}(\theta,\phi)
{\bf X}_{lm}^{*(k)}(\theta^\prime,\phi^\prime)
\nonumber \hspace{10mm} \\ \times
\int \bar r^2d\bar r\,
\gamma_{lm,a}^{ij}(r,\bar r) V_a(\bar r)
\gamma_{lm,b}^{jk}(\bar r,r^\prime) V_b(r^\prime).
\hspace{10mm}
\end{eqnarray}
We observe the separation of the angular coordinates in this form,
which is attributable to the spherical symmetry of the configuration
of concentric sphere geometry. 
Using this feature as a cornerstone, we expand the
logarithm as a series. In each term of the series
the angular terms separate after repeated
use of orthogonality relations for the 
spherical vector eigenfunctions. 
This allows for the separation of the
angular coordinates completely and in conjunction with the trace
in the equation the angular coordinates drop out of the equation,
leaving a sum over $l$ and a factor of $(2l+1)$ from the sum over $m$. 
The leftover series involves integrals in radial coordinates,
which, remarkably, allows for the series to be resummed.
These manipulations, which are mostly formal rearrangement
of integrals, are crucial part of the calculation and
leads to the expression
\begin{eqnarray}
E_{12}(a,b) &=& \frac{1}{2} \int_{-\infty}^\infty
\frac{d\zeta}{2\pi}\sum_{l=0}^\infty (2l+1) 
\nonumber \\ && \times 
\ln \big[1-K_l^E(a,b) \big] \big[1-K_l^H(a,b) \big],
\hspace{5mm}
\end{eqnarray} 
where
\begin{eqnarray}
K_l^E(a,b) &=&  \zeta^2
(\varepsilon_1-\varepsilon_3) (\varepsilon_2-\varepsilon_3)
\int_0^a r^2dr \int_b^\infty {r^\prime}^2dr^\prime
\nonumber \\ && \times 
g_{l,a}^E(r^\prime,r) g_{l,b}^E(r,r^\prime)
\hspace{5mm}
\label{KEabe1}
\end{eqnarray}
and
\begin{eqnarray}
K_l^H(a,b) = 
\left( \frac{1}{\varepsilon_3} -\frac{1}{\varepsilon_1} \right)
\left( \frac{1}{\varepsilon_3} -\frac{1}{\varepsilon_2} \right)
\int_0^a r^2dr \int_b^\infty {r^\prime}^2dr^\prime
\nonumber \\ \times
\text{tr} \left[ \begin{array}{cc}
D^\prime D g_{l,a}^H(r^\prime,r) 
& D^\prime ik_\perp g_{l,a}^H(r^\prime,r) \\
-ik_\perp^\prime D g_{l,a}^H(r^\prime,r)
& k_\perp k_\perp^\prime g_{l,a}^H(r^\prime,r)
\end{array} \right] \hspace{10mm}
\nonumber \\ \times
\left[ \begin{array}{cc}
D^\prime D g_{l,b}^H(r,r^\prime)
& D^\prime ik_\perp g_{l,b}^H(r,r^\prime) \\
-ik_\perp^\prime D g_{l,b}^H(r,r^\prime) 
& k_\perp k_\perp^\prime g_{l,b}^H(r,r^\prime)
\end{array} \right].
\hspace{10mm}
\end{eqnarray}
The integration limits on the coordinate $r$ spans
the inner spherical region from 0 to $a$, and the integration
limits on the radial coordinate $r^\prime$ spans
the outer spherical region beyond $b$, and, together,
they span disjoint regions in space. This 
segregation of variables avoids
ultraviolet divergences in the energy associated
with $r\to r^\prime$.

Evaluating the expression in Eq.\,(\ref{KEabe1})
after substituting the solutions for Green's functions from
Eqs.\,(\ref{gfsolrrp}) we observe the factorization
\begin{equation}
K_l^E(a,b) = r_{31}^E(a) r_{32}^E(b),
\end{equation}
where $r_{ij}^E$ are the scattering coefficients
for the transverse electric mode of an electromagnetic wave 
incident on interfaces $a$ or $b$. The transverse electric 
scattering coefficients at the two interfaces
can be expressed in the form
\begin{subequations}
\begin{eqnarray}
r_{31}^E(a) &=& \frac{1}{a^2} \frac{(\zeta_1^2-\zeta_3^2)
\int_0^a r^2dr\, \text{i}_l(\zeta_1r) \text{i}_l(\zeta_3r)}
{ \left[ \zeta_3 \, \text{i}_l(\zeta_1a) \bar{\text{k}}_l(\zeta_3a)
-\zeta_1 \, \bar{\text{i}}_l(\zeta_1a) \text{k}_l(\zeta_3a) \right]}, \\
r_{32}^E(b) &=& \frac{1}{b^2} \frac{(\zeta_2^2-\zeta_3^2)
\int_b^\infty r^2dr \, \text{k}_l(\zeta_2r) \text{k}_l(\zeta_3r)}
{ \left[ \zeta_2 \, \text{i}_l(\zeta_2b) \bar{\text{k}}_l(\zeta_2b)
-\zeta_3 \, \bar{\text{i}}_l(\zeta_3b) \text{k}_l(\zeta_3b) \right]}.
\hspace{13mm}
\end{eqnarray}
\end{subequations}
The integrals appearing in the numerators of the transverse electric 
scattering coefficients can be evaluated using the
identities~\cite{Prudnikov:1990,Mathematica}
\begin{subequations}
\begin{eqnarray}
\int_0^x y^2dy\, \text{i}_l(py) \text{i}_l(qy)
\nonumber \hspace{55mm} \\ \hspace{5mm}
=- \frac{x^2}{(p^2-q^2)} \Big[ q \text{i}_l(px) \bar{\text{i}}_l(qx)
-p \bar{\text{i}}_l(px) \text{i}_l(qx) \Big], \hspace{12mm} \\
\int_x^\infty y^2dy\, \text{k}_l(py) \text{k}_l(qy) 
\nonumber \hspace{52mm} \\ \hspace{5mm}
=\frac{x^2}{(p^2-q^2)} \Big[ q \text{k}_l(px) \bar{\text{k}}_l(qx)
-p \bar{\text{k}}_l(px) \text{k}_l(qx) \Big], \hspace{12mm}
\end{eqnarray}
\end{subequations}
which immediately leads to the expression for
the transverse electric scattering coefficients in 
Eqs.\,(\ref{refcoe-sp-def}).
The contribution from the transverse magnetic mode can be similarly
factorized into
\begin{equation}
K_l^H(a,b) = r_{31}^H(a) r_{32}^H(b),
\end{equation}
where $r_{ij}^H$ are the scattering coefficients
for the transverse magnetic mode of an electromagnetic wave
incident on interfaces $a$ or $b$. The transverse magnetic scattering
coefficients can be expressed as
\begin{subequations}
\begin{eqnarray}
r_{31}^H(a) = \frac{1}{a^2} (\zeta_1^2-\zeta_3^2)
\nonumber \hspace{55mm}\\ \times 
\frac{ \int_0^a r^2dr\, \left[
\bar{\text{i}}_l(\zeta_1r) \bar{\text{i}}_l(\zeta_3r)
+ \frac{l(l+1)}{r^2 \zeta_1 \zeta_3}
\text{i}_l(\zeta_1r) \text{i}_l(\zeta_3r) \right]}
{ \left[ \zeta_1 \, \text{i}_l(\zeta_1a) \bar{\text{k}}_l(\zeta_3a)
-\zeta_3 \, \bar{\text{i}}_l(\zeta_1a) \text{k}_l(\zeta_3a) \right]}, 
\hspace{12mm}\\
r_{32}^H(b) = \frac{1}{b^2} (\zeta_2^2-\zeta_3^2)
\nonumber \hspace{56mm}\\ \times
\frac{ \int_b^\infty r^2dr \, \left[
\bar{\text{k}}_l(\zeta_2r) \bar{\text{k}}_l(\zeta_3r)
+ \frac{l(l+1)}{r^2 \zeta_2 \zeta_3}
\text{k}_l(\zeta_2r) \text{k}_l(\zeta_3r) \right]}
{ \left[ \zeta_3 \, \text{i}_l(\zeta_3a) \bar{\text{k}}_l(\zeta_2a)
-\zeta_2 \, \bar{\text{i}}_l(\zeta_3a) \text{k}_l(\zeta_2a) \right]},
\hspace{12mm}
\end{eqnarray}
\end{subequations}
where the integrals appearing in the numerators
can be evaluated using the identities~\cite{Prudnikov:1990,Mathematica}
\begin{subequations}
\begin{eqnarray}
\int_0^x y^2dy\, \left[ \bar{\text{i}}_l(py) \bar{\text{i}}_l(qy)
+\frac{l(l+1)}{y^2 pq} \text{i}_l(py) \text{i}_l(qy) \right]
\nonumber \hspace{20mm} \\ \hspace{3mm}
=\frac{x^2}{(p^2-q^2)} \Big[ p \text{i}_l(px) \bar{\text{i}}_l(qx)
-q \bar{\text{i}}_l(px) \text{i}_l(qx) \Big], \hspace{17mm} \\
\int_x^\infty y^2dy\, \left[ \bar{\text{k}}_l(py) \bar{\text{k}}_l(qy)
+\frac{l(l+1)}{y^2 pq} \text{k}_l(py) \text{k}_l(qy) \right]
\nonumber \hspace{15mm} \\ \hspace{3mm}
=-\frac{x^2}{(p^2-q^2)} \Big[ p \text{k}_l(px) \bar{\text{k}}_l(qx)
-q \bar{\text{k}}_l(px) \text{k}_l(qx) \Big], \hspace{12mm}
\end{eqnarray}
\end{subequations}
which leads leads to the expression for
the transverse magnetic scattering coefficients in
Eqs.\,(\ref{refcoe-sp-def}).
Thus, we obtain the expression for the Lizshitz interaction energy
in terms of scattering coefficients to be
\begin{eqnarray}
E_{12}(a,b) = \frac{1}{2} \int_{-\infty}^\infty
\frac{d\zeta}{2\pi}\sum_{l=0}^\infty (2l+1)
\nonumber \hspace{25mm} \\ \times
\ln \big[1- r_{31}^E(a) r_{32}^E(b) \big] 
\big[1-r_{31}^H(a) r_{32}^H(b) \big].
\hspace{5mm}
\end{eqnarray}
This expression for Lifshitz interaction energy is for zero temperature.
The interaction energy for nonzero temperature
in Eq.\,(\ref{enSpL}) is obtained from the above expression
by the replacement
\begin{equation}
\frac{1}{2} \int_{-\infty}^\infty \frac{d\zeta}{2\pi}
\to \frac{\hbar c}{2\pi a_0} {\sum_{n=0}^\infty}{^\prime}.
\end{equation}

%-----------------------------------------------------

\end{document}